\def \yskip{\penalty-50\vskip3pt plus 3pt minus 2pt}
\def \reference{\par \yskip \noindent \hangindent .4in \hangafter 1}
\def \abc#1#2#3#4 {\reference#1, {\sl#2}, {\bf#3}, #4}
\def \blank {\lower 5pt\hbox to 0.75in{\hrulefill}}

\def \cm{~\rm{cm}}
\def \s{~\rm{s}}
\def \km{~\rm{km}}

\def \K{~\rm{K}}

\def \AU{~\rm{AU}}
\def \ergs{~\rm{ergs}}
\def \yrs{~\rm{yrs}}
\def \yr{~\rm{yr}}

\def \kpc{~\rm{kpc}}
\def \lesssim{\mathrel{<\kern-1.0em\lower0.9ex\hbox{$\sim$}}}
\def \gtrsim{\mathrel{>\kern-1.0em\lower0.9ex\hbox{$\sim$}}}

\documentstyle[11pt,aasms,tighten,flushrt]{article}
\begin{document}
%\normalsize
\small

\setcounter{page}{1}
% simwin7.tex (12 Oct. 1999: WITH CORRECTIONS INSIDE, Sub. To ApJ)

\begin{center}
\bf
THE FORMATION OF VERY NARROW WAIST \\
BIPOLAR PLANETARY NEBULAE
\end{center}
%\vspace*{2.0cm}

\begin{center}
Noam Soker\\
Department of Physics, University of Haifa at Oranim\\
Oranim, Tivon 36006, ISRAEL \\
soker@physics.technion.ac.il \\
and\\
Saul Rappaport \\
Physics Department, MIT \\
Cambridge, MA 02139 \\
sar@space.mit.edu
\end{center}

%\clearpage
%$$
%$$

\begin{center}
\bf ABSTRACT
\end{center}

  We discuss the interaction of the slow wind blown by an asymptotic giant
branch (AGB) star with a collimated fast wind (CFW) blown by its main
sequence or white dwarf companion, at orbital separations in the range of
several$~ \times \AU \lesssim a \lesssim 200 \AU$.
 The CFW results from accretion of the AGB wind into an accretion disk
around the companion.  The fast wind is collimated by the accretion disk.
 We argue that such systems are the progenitors of bipolar planetary
nebulae and bipolar symbiotic nebulae with a very narrow equatorial
waist between the two polar lobes.
 The CFW wind will form two lobes along the symmetry axis,
and will further compress the slow wind near the equatorial plane,
leading to the formation of a dense slowly expanding ring.
 Therefore, contrary to the common claim that a dense equatorial ring
collimates the bipolar flow, we argue that in the progenitors of very
narrow waist bipolar planetary nebulae, the CFW, through its interaction
with the slow wind, forms the dense equatorial ring.
 Only later in the evolution, and after the CFW and slow wind cease,
the mass losing star leaves the AGB and blows a {\it second}, more
spherical, fast wind.
 At this stage the flow structure becomes the one that is
commonly assumed for bipolar planetary nebulae,
i.e., collimation of the fast wind by the dense equatorial material.
 However, this results in the broadening of the waist in the equatorial
plane, and cannot by itself account for the presence of very narrow waists
or jets.
 We conduct a population synthesis study of the formation of planetary
nebulae in wide binary systems which quantitatively supports
the proposed model.  The population synthesis code follows the evolution of
both stars and their arbitrarily eccentric orbit, including mass loss via
stellar winds, for $5 \times 10^4$ primordial binaries.
 We show the  number of expected systems that blow a CFW is in accord
with the number found from observations, to within the many
uncertainties involved.
 Overall, we find that $\sim 5 \%$ of all planetary nebulae are bipolars
with very narrow waists.
  Our population synthesis not only supports the CFW model, but
more generally supports the binary model for the formation of
bipolar planetary nebulae.

{\it Subject headings:} planetary nebulae: general
$-$ stars: binaries: close
$-$ stars: AGB and post-AGB
$-$ stars: mass loss
$-$ ISM: general

%\clearpage

% ======================================================================
\section{INTRODUCTION}
% ======================================================================

  Bipolar (also called ``bilobal'' and ``butterfly'') planetary nebulae
(PNs) are defined (Schwarz, Corradi \& Stanghellini 1992) as
axially symmetric PNs having two lobes with an `equatorial'
waist between them.
 Corradi \& Schwarz (1995) estimate that $\sim 11 \%$ of all PNs
are bipolars.
 The most commonly assumed flow structure for the formation of the
bipolar structures, in PNs and other objects, e.g., Luminous Blue Variables,
is the ``Generalized Wind Blown Bubble'' (GWBB; see review by Frank 1999).
 In the GWBB flow structure, a fast tenuous wind is blown into a previously
ejected slow wind. The slow wind is assumed to have a higher density
near the equatorial plane.
  The higher slow-wind density near the equatorial plane forces the fast
wind to blow a prolate nebula, with the major axis along the symmetry axis
(e.g., Soker \& Livio 1989; Frank \& Mellema 1994; Mellema \& Frank 1995;
Mellema 1995; Dwarkadas, Chevalier, \& Blondin 1996).
 When the equatorial to polar density ratio is very high, a bipolar nebula
is formed.
 The numerical simulations in the works cited above use axisymmetrical
slow winds and spherical fast winds, and do not form bipolar PNs with
very narrow waists, but rather form elliptical PNs, or bipolar with
wide waists.
 Some other problems of the GWBB model are mentioned by Balick (2000). 
  It seems that in order to form bipolar PNs with very narrow waists,
a collimated fast wind (CFW) is required.
 Such simulations were performed by Frank, Ryu, \& Davidson (1998),
who indeed got very narrow waists in their simulated nebulae.

 To collimate a fast wind, a dense gas is required to be present in the
equatorial plane very close to the star that blows the wind, e.g.,
an accretion disk.
 This dense material is supplied by a slow wind from the companion star.
 Therefore, it seems that in order to form very narrow waist bipolar PNs,
the fast wind and the slow winds are being blown simultaneously,
at least during part of the evolution.
In the present paper we examine some conditions for the formation of
the two winds, and the interaction between the two winds.
 We further assume that the slow wind is blown by an asymptotic giant
branch (AGB) star,  while the fast wind is blown by a main sequence or
a WD companion (Morris 1987).
 A later fast wind will be blown by the mass losing star after it
leaves the AGB.
 The idea that detached binary systems can expel winds with equatorial
mass concentration, and hence lead to the formation of bipolar PNs, goes back
to Livio, Salzman, \& Shaviv (1979) and Morris (1981).
 The interaction between the binary components was studied numerically
by Mastrodemos \& Morris (1998, 1999; hereafter MM98 and MM99).
 These works, though, did not consider any CFW.
 Morris (1987; 1980) proposed and studied the process
by which the giant star blows a slow wind, part of which is
accreted by the companion, which blows a CFW.
 Support for this model comes from bipolar symbiotic nebulae
(e.g., He 2-104; see image in Corradi \& Schwarz 1995).
 The connection between bipolar PNs and symbiotic nebulae was pointed out
by Morris (1990) and Corradi and Schwarz in a series of papers
(e.g., Corradi 1995; Corradi \& Schwarz 1995; Schwarz \& Corradi 1992).
 In a recent paper Lee \& Park (1999) argue that an accretion disk is
present around the hot companion of the symbiotic star RR Telescopii.
 Another object which suggests that the bipolar structure forms before the
post AGB star blows a fast wind is the bipolar nebula OH 231.8+4.2
(Kastner {\it et al.} 1998; also called OH 0739-14).
This nebula has a regular Mira variable in its center,
which seems to have a blue companion (Cohen {\it et al.} 1985).
 Cohen {\it et al.} (1985) suggest that the bipolar structure is due to
mass loss from the binary system, and they further point to the
connection of this system to both symbiotic stars and bipolar PNs.

 In a recent paper, Soker (1998a) lists properties of bipolar PNs,
and discusses each in the context of binary versus single star models.
His conclusion was that binary models can explain all these properties,
while models based on single star evolution have a hard time explaining
several of these properties.
 Soker further argues that most bipolar PNs are formed from binary
stellar systems which avoid a common envelope phase
during a substantial fraction of the interaction process.
 The positive correlation of bipolar planetary nebulae with massive
progenitors, $M \gtrsim 2 M_\odot$, is attributed by Soker (1998a) to the
larger ratio of red giant branch (RGB) to AGB radii which low mass
stars attain, compared with massive stars.
 These larger radii on the RGB cause most binary stellar companions,
which potentially could have helped form bipolar PNs if the primary had been
on the AGB, to interact with the low mass primaries while
they are still on the RGB.
Even a weak tidal interaction which spins up the primary RGB star by only
a modest amount, can result in a much higher mass loss rate, which
leaves the star with an insufficient envelope to ascend the upper AGB.
  That low-mass RGB stars can lose almost their entire envelope is
evident from the distribution of horizontal branch stars in many
globular clusters (e.g., Ferraro {\it et al.} 1998, and references therein).
 The mechanism behind this enhanced mass loss is not known, but it
seems to result from faster rotation or tidal interaction
(e.g., Ferraro {\it et al.} 1998, and references therein).
 This is further examined in $\S 4$.

  The three symbiotic nebulae presented by Corradi \& Schwarz (1995),
show both very-narrow waists, and a ``Crab''-like  structure, that is,
the images show two ``arms'' on each side of the equatorial plane.
  Based on that, we extend our definition of very-narrow waist PNs to
include those which show clear ``Crab''-like images; we predict that future
higher resolution images of the inner regions of these PNs may show the
structures of very-narrow waists (unless the second fast wind destroys
the narrow waist).
 Images of 43 bipolar PNs are presented by Corradi \& Schwarz (1995).
 We list these PNs in Table 1 (the first column is the common name and the
second column gives the PN G name, according to the galactic coordinates),
and indicate in the third column those which we refer to as ``very-narrow
waist PNs''.
 A question mark means we cannot decide on a classification based on
the image we have.
The fourth column gives the maximum expansion velocity, also taken from
Corradi \& Schwarz (1995).
The fifth column indicates those for which we see point symmetry,
and the last column indicates those for which we can
see a clear deviation from axisymmetry.
 The point symmetry may hint at precession of a CFW due to a binary
companion which exerts torque on a tilted disk
(an alternative is a precession due to a disk instability
Livio \& Pringle 1996),
while deviation from axisymmetry may hint at the presence
of a companion in an eccentric orbit (Soker, Rappaport \& Harpaz 1998).
 We note that other very-narrow waist bipolar PNs exist, e.g.,
the 12 images presented by Sahai \& Trauger (1998) contain two
very-narrow waist bipolar PNs:  HB 12 and He2-104.
However, we will concentrate on the list of Corradi \& Schwarz
(1995), on which they based their estimate of the fraction of
bipolar PNs among all PNs, of $\sim 11\%$.

 In $\S 2$ we present the proposed scenario for the formation of bipolar
PNs with very narrow waists, and in $\S 3$ we discuss the interaction
of the CFW with a slow wind.
 In $\S 4$ we describe our population synthesis and evolution calculations,
which we use to check quantitatively the binary model for bipolar PNs in
general, and the proposed CFW model as a subclass of bipolar PN progenitors
in particular.
 The results of the population synthesis and related discussions are
given in $\S 5$, while our main findings are summarized in $\S 6$.

% ======================================================================
\section{THE FORMATION OF A COLLIMATED FAST WIND (CFW)}
% ======================================================================

 Three processes for the formation of high density mass flow in the
equatorial plane of AGB stars are mentioned in the literature.
 In the present paper we suggest a fourth one.
 The first process is due to the rotation of the mass losing star.
Fast rotation $v_{\rm spin} \gtrsim 0.3 ~v_{\rm Kep}$,
where $v_{\rm spin}$ and $v_{\rm Kep}$ are the equatorial rotation
and the break-up velocities (i.e., the maximum rotational speed, which is
equal to the Keplerian velocity on the stellar equator) of the star,
respectively, can lead to
dense equatorial flow due to dynamical effects
(Garcia-Segura {\it et al.} 1999 and references therein).
 Slower rotation velocities can lead to higher mass loss rate in the
equatorial plane as well, e.g., via the activity of a stellar dynamo.
 This process is significant in the shaping of bipolar PNs only
if the envelope was spun up via a common envelope or
tidal interaction (Soker 1998b).
 We suggest that this process by itself will not form very narrow waist
PNs, but may form bipolar PNs, as is the case with NGC 2346,
a bipolar PN whose progenitor went through a common envelope phase
(Bond \& Livio 1990).

 The two other processes, and the one proposed here, are due to
a close stellar companion.
 The second process is the gravitational focusing of the primary's wind by
the companion, and the third is due to the orbital motion of the
primary around the center of mass.
 These processes have been studied in detail by MM99, where earlier
references can also be found.
 They find that a close companion can form a very high density equatorial
flow.
The equatorial to polar density contrast depends on the terminal wind
velocity and its acceleration distance away from the star,
the masses of the two stars, and the orbital separation $a$.
 The effect of the orbital motion was estimated by Soker (1994).
 The velocity of the mass losing star around the center of mass is given by
\begin{eqnarray}
v_1 =
9.4
\frac {M_2}{M_\odot}
\left( \frac {M_1+M_2}{M_\odot} \right) ^{-1/2}
\left( \frac {a}{10 \AU} \right) ^{-1/2}
\km \s^{-1},
\end{eqnarray}
where $M_1$ and $M_2$ are the masses of the mass losing star
and its companion, respectively.
 This velocity should be compared with the wind velocity.
 If the acceleration zone of the wind is of the order of the
orbital separation, then the process is more complicated (MM99).
The velocity of the slow wind blown on the upper AGB
is $v_s \simeq 10 \km s^{-1}$.
 If $v_1 \sim v_s$, then a strong concentration toward the
equatorial plane will occur due the orbital motion alone
(Soker 1994).
 For $v_1>v_s$ the concentration will be stronger and no mass will be found
along the symmetry axis away from the binary system.
  The implications of these effects for the scenario proposed in this
work are discussed in the next section.

 In the present paper we propose that a wind blown by the {\it companion}
to the AGB star can further concentrate the flow toward the
equatorial plane, as well as causing other morphological features.
 We now outline the proposed scenario, and in the next section
we present the basic flow structure.
Some relevant time scales, e.g., tidal synchronization, can be found
in Soker (1998a), and we will not repeat their derivation here.
 As the mass losing star evolves along the AGB the ratio of its
radius to the orbital separation increases (as long as the mass
loss rate is not too high).
 When this ratio becomes $R_1 /a \gtrsim 0.1 (M_2/M_1)^{-1/3}$,
tidal interactions become important, and the secondary spins-up the
AGB star's envelope to synchronization.
 In the above expression we assume $M_2<M_1$.
 For AGB stars this means that the orbital separation should be
in the range of $\sim 1 \AU - 20 \AU$.
 The upper limit is for massive stars which are near the tip of the AGB.
 Since stars on the AGB reach radii of $\gtrsim 1 \AU$, for orbital
separations of $\lesssim 5-10 \AU$ circularization will be achieved
as well as synchronization (Soker 1998a), and an intensive mass
transfer, via captured enhanced equatorial wind, will occur (MM98).
 If such an increase results in the formation of a disk around the
secondary star (MM98) and a collimated outflow
(Morris 1987; 1990) as we assume here, then the kinetic energy outflow
of the CFW can exceed that of the slow wind.
 This CFW will shape the slow wind into a very narrow waist bipolar PN.
 Later in the evolution, the star leaves the AGB, and blows
a {\it second} fast wind, but now, at least in certain directions,
there is no slow wind anymore.
 We argue that the formation of very-narrow waist bipolar PNs
requires the formation of a CFW.
 The condition $v_1 \gtrsim v_s$ discussed above can also lead to
a dense slow equatorial flow (see MM99), but we find that this condition
implies the formation of a CFW.
  In the next section we argue that the CFW interaction with the
slow wind will lead to the formation of a slowly expanding dense
ring in the equatorial plane.

  It is generally agreed that a disk is a necessary condition for the
formation of jets emanating from compact objects
(for a recent review see Livio 2000).
 In the model proposed in the present paper, the CFW can have a wide
opening angle, and is not limited to a narrow jet.
 We therefore do not limit the properties of the disk, e.g., that it
be extended, but rather only require that an accretion disk is
formed around the accreting star.
  This condition reads $j_a > j_2$, where $j_a$ is the specific
angular momentum of the accreted material, and $j_2=(G M_2 R_2)^{1/2}$
is the specific angular momentum of a particle in a Keplerian orbit at
the equator of the accreting star of radius $R_2$.
 We perform the calculations here for the case where the initial
secondary star is the accretor.
 The same expressions hold when the
accretor is the WD remnant of the initial primary star, for which
the mass and radius are dramatically different.
 For accretion from a wind, the net specific angular momentum of the
material entering the Bondi-Hoyle accretion cylinder, i.e., having
impact parameter $b<R_a = 2 G M_2 /v_r^2$, where $v_r$ is the
relative velocity of the wind and the accretor, is (Wang 1981)
$j_{BH} = 0.5 (2 \pi / P_o) R_a^2$, where $P_o$ is the orbital period.
  Livio {\it et al.} (1986; see also Ruffert 1999) find that the actual
accreted specific angular momentum for high Mach number flows are
$j_a = \eta j_{BH}$, where $\eta \sim 0.1$ and $\eta \sim 0.3$
for isothermal and adiabatic flows, respectively.
 The relative velocity is $v_r^2 \simeq v_s^2 + v_o^2$, where $v_s$
is the (slow) wind velocity at the location of the accreting star, and $v_o$
is the relative orbital velocities of the two stars.
 Considering that the orbital velocity goes as $a^{-1/2}$,
where $a$ is the orbital separation, and the wind velocity increases
with radial distance close to the primary (MM99), we simply approximate
$v_r$ to be a constant equal to $15 \km \s^{-1}$.
 Substituting typical values for WD accretor and the mass-losing
terminal AGB star we find the following condition for the
formation of a disk
\begin{eqnarray}
1< \frac {j_a}{j_2} = 16
\left( \frac {\eta}{0.2} \right)
\left( \frac {M_1+M_2}{1.2 M_\odot} \right)
\left( \frac {M_2}{0.6 M_\odot} \right)^{3/2}
\left( \frac {R_2}{0.01 R_\odot} \right)^{-1/2}
\left( \frac {a}{10 \AU} \right)^{-3/2}
\left( \frac {v_r}{15 \km \s^{-1}} \right)^{-4} .
\end{eqnarray}
  From this last equation,
it turns out that a disk around a WD can be formed up to an orbital
separation of $a \sim 60 \AU$, while around main sequence
stars with $R_2 \sim R_\odot$, disk can formed up to orbital separations
of $a \sim 15 \AU$ for $M_2 \sim M_\odot$, or to larger orbital separations
for more massive main sequence stars.
 For $M_1=1.5 ~M_\odot$ and $M_2=1 ~M_\odot$, as in the standard models
of MM99, and the other parameters as in the equation above,
we find that a disk will be formed to a distance of  $a \simeq 37 \AU$.
Considering the many uncertainties, especially in the
effective value of $v_r$ and $\eta$, this is quite close to the finding of
MM99 of $a \simeq 24 \AU$ for their slow wind case.
 Put another way, a value of $\eta=0.1$ matches better the results of
MM99.

Another plausible condition for the formation of a CFW is that the
accretion rate should be above a certain limit $\dot M_{\rm crit}$,
which we take as $10^{-7} M_\odot \yr^{-1}$ for accretion onto a
main sequence star and $10^{-8} M_\odot \yr^{-1}$ for accretion
onto a WD. For a discussion of these limiting values of $\dot M$ see
$\S 3.1$  and $\S 4.2$)
 The accretion rate depends on several factors (MM99), e.g.,
the accretor mass, the acceleration zone of the slow wind,
concentration toward the equatorial plan, and synchronization of
the mass losing star.
 We neglect most of these effects, and simply take the accretion
rate to be $\dot M_2 = (R_a/2 a)^2 \vert \dot M_1 \vert$.
Substituting the relevant parameters during the superwind phase, i.e.,
the high mass loss rate from an AGB star, we get
\begin{eqnarray}
\dot M_2 \simeq
5 \times 10^{-6}
\left( \frac {M_2}{0.6 M_\odot} \right)^{2}
\left( \frac {v_r}{15 \km \s^{-1}} \right)^{-4}
\left( \frac {a}{10 \AU} \right)^{-2}
\left( \frac {\vert \dot M_1 \vert }{10^{-4} M_\odot \yr^{-1}} \right)
M_\odot \yr^{-1}.
\end{eqnarray}
 This condition is weaker than the one on the specific angular momentum,
if indeed the AGB star blows a significant superwind.
 A more detailed expression for the accretion rate is given by
Morris (1990) and Han, Podsiadlowski \& Eggleton (1995).
  Han {\it et al.} take the condition for the formation of what they
term a bipolar PN, to be $\dot M_2 / \vert \dot M_1 \vert \gtrsim 0.1$.
 However, by bipolar PNs they refer to all highly asymmetrical PNs,
and not in particular to those having waists.

 The formation of a dense equatorial wind due to the orbital motion requires
$v_1>v_s$. Using equation (1) for $v_1$, this condition reads
\begin{eqnarray}
\frac {M_2}{0.6 M_\odot}
\left( \frac {M_1+M_2}{1.2 M_\odot} \right) ^{-1/2}
\left( \frac {a}{10 \AU} \right) ^{-1/2}
\left( \frac {v_s}{10 \km \s^{-1}} \right)^{-1}
\gtrsim 2.
\end{eqnarray}
 For the standard model of MM99 ($M_1=1.5 M_\odot$; $M_2=1 M_\odot$)
this condition requires $ a \lesssim 3 \AU$.
However, because the wind has an extended acceleration zone,
its velocity at the center of mass is much below the terminal velocity.
 This means the effective value of $v_s$ will be lower, and the condition
$v_1 > v_s$ can hold up to a distance of $a \sim 10 \AU$.
 This indeed can be observed in the models of MM99.
Comparing conditions (2) with (4), we see that the formation of
a dense equatorial flow due to the orbital motion implies the formation
of a CFW, in almost all relevant cases.

%\clearpage
% ======================================================================
\section{THE INTERACTION OF THE CFW WITH THE SLOW WIND}
% ======================================================================
% ======================================================================
\subsection{General Properties}
% ======================================================================

 The CFW blown by the secondary star (or the WD remnant of the primary if
the initial secondary is in the terminal AGB phase)
interacts with three types of media:
(i) The slow wind blown at the same time by the AGB star, in a region
located between the two stars.
This is shown schematically in Figures 1a and 2a.
(ii) The slow wind material ejected within a time of less than an orbital
period, and which is located on the other side of the accreting star.
This is drawn schematically on the left hand sides of Figures
1b and 2b.
(iii) With material expelled earlier than one orbital period before
the present time.
This material includes shells formed from earlier interactions of
the CFW with the slow wind.
Such shells are drawn on Figures 1b and 2b (e.g., region F).
 Further out in the nebula, these shells interact with slow wind material
blown before the onset of the CFW.
 In Figures 1 and 2 we show schematically the flow structure in a
plane perpendicular to the equatorial plane, and
containing the two stars.
 Figures 1a and 2a show the inner regions of Figures 1b and 2b, respectively.

 The interaction of the CFW with the slow wind in the region
between the two stars can result in two qualitatively different flow
structures.
 These are drawn schematically on Figure 1 and 2, respectively.
 In the first, the ``strong CFW case'', the deflection angle $\theta_d$
is smaller than the collimation angle $\theta_c$:
$\theta_d < \theta_c$ (both angles are defined in Figs. 1 and 2),
while in the second, the ``weak CFW case'', $\theta_d > \theta_c$.
 The angle $\theta_d$ depends on the angle of collimation
$\theta_c$, on the ratio of the momentum flux in the
slow wind to that in the CFW, the deviation of the slow wind from spherical
shape, and on the flow of the accreted slow wind near the secondary.
 In the first type of flow (Fig. 1) there is an avoidance region for
the slow wind near the symmetry axis (region E on Fig. 1b),
while in the second type (Fig. 2) there is an avoidance region for
the CFW there (region E on fig. 2b).
 In this later case, dense and slowly expanding material
will flow along the symmetry axis.
 The image of the PN NGC 6302 (Hua, Dopita, \& Martinis 1998),
and of the symbiotic nebulae BI Cru show dense material along the
symmetry axis.
 This slow material may block the radiation from the central star,
leading to the formation of what is termed ``searchlight beams'',
e.g, the Egg Nebula.
 In general, in close binary systems at early stages when the CFW is weak,
the flow depicted in Figure 2 will commence, while later, the flow
depicted in Figure 1 will take over.
Therefore, we might find slow material along the symmetry axis away from
the center.
If this second stage does not occur, then we might get the
``searchlight beams'' as the primary leaves the AGB and lights up the
nebula.
 In wider binaries ($a \gtrsim 20 \AU$) which result in elliptical PNs,
only the weak CFW case occurs (unless the CFW is well collimated, i.e.,
a jet; see next subsection).

 If the orbital separation and secondary mass are such that
the primary's orbital velocity (eq. 1) is $v_1 \gtrsim v_s$,
then very little slow wind will be blown along the symmetry axis (MM99).
 Here $v_s$ should be taken as the wind velocity at the center of mass,
rather than the terminal velocity.
However, if the physical circumstances produce a weak CFW, then
slow wind material which is shocked when colliding with the CFW,
will expand into region E of Figure 2b.
 Therefore, region E will be filled with slow wind material, even
though no material which leaves the primary has velocity perpendicular
to the orbital plane
(due to the primary's large orbital velocity).

 Let us describe the periodic flow of material along a specific direction.
 Let it be the direction to the left of the accretor in figure 1b and 2b.
 As the CFW collides with the slow wind on the side of the
secondary (left sides of figures 1 and 2), it accelerates it
and ``cleans'' the side of the secondary from the slow wind.
 A quarter of an orbital period later, $0.25 P$, the slow wind material
which is blown in the direction of motion of the primary, at a velocity of
$v_s+v_1$,  will flow in that direction (left on Figures 1 and 2).
  After one orbital period, this wind will have reached a distance of
\begin{eqnarray}
L \simeq  0.75 P (v_s+v_1) \sim 10 a,
\end{eqnarray}
where in the second equality we took $v_1 \simeq v_s/2$ and assume
an equal mass binary system in calculating $P$.
 After a time of $3/4 P$ material leaving the primary in the opposite
direction of the primary's orbital motion will flow into this region
(the region that was initially on the secondary side, i.e., left on
Figures 1 and 2). 
 Its velocity in the center of mass rest frame is $v_s-v_1$.
 Therefore, the slow wind material will fill the region inward to
$L$, with increasing radial velocity with distance from the binary system.
 The slow wind material which was accelerated by the CFW
in the previous orbital passage of the secondary
(region B in the figures), will be at a distance $r \simeq 2-5 L$
(see below).

 The motion of the fast wind through the dense, slowly expanding
slow wind, resembles in many ways the propagation of supersonic
jets through a dense medium.
 The bow shocks running ahead of a jet and through the ambient medium,
accelerates the ambient medium.
 Several jet diameters behind the head, though, the shocked ambient
medium flows perpendicular to the jet's propagation direction, or even
backward (e.g., Chernin {\it et al.} 1994).
 This flow compresses the ambient medium, and sweeps up a dense
shroud of ambient gas (Chernin {\it et al.} 1994).
 The shroud on the equatorial side, we argue, will form a dense region
near the equatorial plane.
 A dense flow was present before, due to the aspherical mass loss by
the primary and due to the binary interaction
(e.g., Livio {\it et al.} 1979; Morris 1981; MM98),
but the CFW further concentrates it toward the equatorial plane, and forms
the slowly expanding equatorial ring.
 Therefore, it is the CFW that forms the high density equatorial
flow, contrary to the common claim that this equatorial ring
collimates the fast outflow.
 This ring will collimate the later second fast wind, blown by
the post AGB star itself, $\gtrsim 10^3$ years after it leaves the AGB,
and after the bipolar structure already exists.

 To calculate the velocity of the accelerated shells depicted in
Figures 1 and 2, we carry out some simple calculations.
 In the flow depicted in Figure 1, no hot bubble of shocked fast wind is
formed near the shell, since the hot shocked gas is ``leaking'' along
the axis (region E of figure 1b).
 Therefore, the shell velocity is determined by momentum balance.
  Assuming a spherical slow wind, the fraction of the slow wind
material that enters the shell is (see Fig. 1)
$\sim (1-\cos \theta_c)$, while
the fraction of the fast wind that accelerates the shell is
$\sim [\cos (\theta_c - \theta_d) - \cos \theta_c]/(1-\cos \theta_c)$.
  The shell's velocity is therefore
\begin{eqnarray}
x \equiv {{v_{\rm shell}} \over {v_s}}  \simeq
1+
{{[\cos (\theta_c - \theta_d)- \sin \theta_c]}\over{(1-\cos \theta_c)^2}}
{{\dot M_f v_f}\over{\dot M_s v_s}},
\end{eqnarray}
where as before, $v_f$ is the velocity of the CFW, $\dot M_f$ the mass
loss rate into the CFW, and $v_s$ the velocity of the slow wind
blown by the primary AGB star.
 Since in this flow the momentum flux of the CFW is larger,
we find for reasonable values $x \gtrsim 2$.
  There are two phases to the acceleration period of the shell.
In the first, the shell is still within the slow wind region (region `C'
in Fig. 1 and 2), and more mass is being added to the shell.
In this phase the shell velocity is more or less constant.
 In the second phase the shell, which moves faster than
the slow wind, leaves region C, and runs into a very low density medium.
In this second part the shell accelerates, and it ends when, due to the
orbital motion, the CFW stops hitting it. 
In the strong CFW case (Fig. 1b), the shell will open up,
forming a crab-like structure.
 Hence, the fast wind will ``slide'' along the shell, losing less of its
radial momentum to the shell.
This reduces somewhat the efficiency of the acceleration process.
 The flow near region `I' is very complicated.
 This is due to the fact that it contains both slow wind material that
was deflected by the strong CFW in the region {\it between} the stars,
and shell material that is entrained by the fast wind streaming on the
edge of the shell or sliding along it. 

  In the flow depicted in Figure 2, the weak CFW case, the momentum flux
from the CFW is smaller than that of the slow wind
(at the same distance from the two stars).
 In this case, the shocked fast wind does not leak out
along the symmetry axis, but rather forms a hot bubble.
This results in a more efficient acceleration of the shell, as in the
energy conserving case in interacting winds in (elliptical) PNs.
 The pressure in the bubble exerts a force on the shell, and accelerates it.
 Here also there are two phases.
 In the first, the bubble is within the slow wind region `C',
and expands at a constant velocity.
 A crude estimate of this velocity can be derived by using the usual
equations of the energy conserving case of interacting winds in PNs
(Volk \& Kwok 1985; Chevalier \& Imamura 1983), taking the non-spherical
geometry into account.
 After the shell breaks out of the slow wind region, the shell accelerates,
and the bubble volume increases substantially.
 In the second phase, when the hot tenuous shocked fast wind acclerates the
dense shell, Rayleigh-Taylor instabilities will develop, breaking the
shell into small clumps.
 This may reduce somewhat the efficiency of the acceleration. 
 For simplicity we assume that most of the fast wind energy is transfered
to thermal energy of the hot bubble, and then is added to the
kinetic energy of the shell.
  This assumption gives for the final shell velocity
\begin{eqnarray}
x^2 \simeq
1+ {{\dot M_f v_f^2} \over {\epsilon \vert \dot M_1 \vert v_s^2}},
\end{eqnarray}
where $\epsilon$ is a geometrical factor, which is actually the
fraction of the slow wind that is captured into the shell. 
 Since for our typical parameters we find that
$\dot M_f \gtrsim 10^{-4} \vert \dot M_1 \vert$, and since
$v_f \sim 100 v_s$, equation (7) indicates that $x \gtrsim 2$.
 So in both types of flows we expect that close to the
binary system the front of the shell will move at a few times the
slow wind velocity, or $v_{\rm shell} \sim 30 \km \s^{-1}$.
 While in the flow depicted in Figure 1, the strong CFW case,
slow wind  material will be accelerated to large velocities in region `I',
($\gg v_{\rm shell}$),  
no such velocities will be attained in the weak CFW case. 
 
  Figures 1 and 2 present the flow structure in a plane.
 In 3 dimensions, the structure will be that of a corkscrew,
close to the binary system.
 As the shells move radially away from the binary system, they expand and
collide with previously formed shells.
  If the secondary ionizes the shell, the sound speed inside the shell will
 be $c_s \sim 10 \km \s^{-1}$, while if the shell is cool, $\sim 1000 \K$,
 the sound speed will be only $c_s \sim 3 \km \s^{-1} \sim 0.2 v_s$.
  As the shells form they are relatively narrow.
  After they cease to be compressed by the CFW (as the binary system
rotates in its orbital motion) they expand at a velocity about equal
to their internal sound speed.
 The time interval between shells along a specific radial direction
is the orbital period $P$.
 Hence the distance beween shells is $D=Pv_{\rm shell} =
2 \pi a v_{\rm shell}/v_{\rm orb}$, where $a$ is the orbital separation
and $v_{\rm orb}$ is the orbital relative velocity of the two stars.
 Therefore, after they move a distance of
$\sim D v_{\rm shell}  / c_s  
=2 \pi a v^2_{\rm shell} /(v_{\rm orb} c_s ) \sim 100-10^3 a$,
adjacent shells will merge. 
 Due to merging of shells, the corkscrew structure will be lost,
and a more axisymmetric structure will take over at a distance of
$10^2-10^3 a$ from the binary system.
 For progenitors of bipolar PNs this distance is $10^{16}-10^{17} \cm$.
 At a large distance from the binary system the shells are slowed down due
to slow wind material that was ejected earlier.
The shells collide, and form a larger shell, the one that is observed
as the ``wall'' of the bipolar structure.
 In a different kind of process, close binary systems
can form equatorial shells due to the gravitational
focusing by the companion to the AGB star (MM99).
 For systems having orbital separtion of $a \lesssim 10 \AU$ the
 gravitational focusing might be more important.

  During the proto-PN phase, the mass losing star has already
left the AGB, and the slow wind and CFW are not active any more.
 The wind from the mass losing star, now a post-AGB star, is faster,
and it smoothes and cleans the region close to the binary system,
where a corkscrew structure was present before.
 Some deviation from perfect axisymmetry can remain though.
 Deviations from axisymmetry are seen in many bipolar PNs, but there
are other possible mechanisms for their formation
(Soker 1994; Soker {\it et al.} 1998).

 Support for our proposed model comes from symbiotic nebulae
(Morris 1990; Corradi \& Schwarz 1995), several proto-PNs,
and supersoft X-ray sources (see below).
Symbiotic nebulae are binary systems where the giant star is blowing a slow
wind, but not yet a fast wind.
 Therefore the CFW, or jets, must be blown by its
compact companion.
 The formation of a jet in an outburst of the symbiotic star CH Cyg
has been reported by Taylor, Seaquist, \& Mattei (1986).
They claim that the mass transfer, which leads to the formation of a
disk around the WD companion, is via Roche lobe overflow.
 We have considered only large orbital separations to avoid computational
complications, although our model can be applied to orbital separations
close enough to allow Roche lobe overflow.
 Another system with a small orbital separation where our model may
apply, is the Red Rectangle (AFGL 915), a proto-PN which has a narrow
waist (Osterbart, Langer, \& Weigelt 1997).
Osterbart {\it et al} (1997) estimate an orbital separation of a few AU,
and they detect a dense disk with an outer radius of $\sim 200 \AU$.

That a dense equatorial mass concentration cannot collimate a bipolar
outflow, is claimed by Bujarrabal, Alcolea, \& Neri (1998) for the
proto-PN M1-92.
 In their observations and analysis of M1-92, Bujarrabal {\it et al.}
(1998) find that the equatorial disk they observe is too large to collimate
the optical jets.
 Other examples of this problem are given by Balick (2000).
Instead, Bujarrabal {\it et al.} (1998) suggest the formation of an
accretion disk in the proto-PN phase by material falling back on the
central star.
Soker \& Livio (1994) already considered the formation of an accretion
disk from material falling back on the central star.
Soker \& Livio (1994), though, argued that this process may occur only
after the fast wind from the central star ceases, and a binary companion
is required for the falling-back material to have enough angular momentum
to form a disk.
 The important point here, though, is that the finding of Bujarrabal
{\it et al.} (1998), supports our claim that the dense material in the
equatorial plane cannot collimate jets to form narrow waists.

	Luminous, galactic supersoft X-ray sources have typical X-ray
luminosities of $\sim 10^{36} - 10^{38} \ergs \s^{-1}$, and characteristic
temperatures of a few$~\times~10^{5} \K$ (see, e.g., Greiner 1996; Kahabka
\& van den Heuvel 1997, and references therein).
 The high temperatures are thought to arise from nuclear burning on
the surface of white dwarfs (van den Heuvel {\it et al.} 1992;
Rappaport, Di Stefano, \& Smith 1994).
Many of these sources have been associated with white dwarfs accreting
at high rates ($\sim 10^{-8} - 10^{-7} M_\odot \yr^{-1}$) from Roche-lobe
filling main-sequence or subgiant companions and from the stellar winds
of giants in symbiotic novae (Greiner 1996 and references therein).
Fast collimated outflows have been observed in three of the supersoft
X-ray sources: RX J0513.9-6951 ($3800 \km \s^{-1}$,
Southwell {\it et al.} 1996; Southwell, Livio, \& Pringle 1997);
RX J0019.8+2156 ($\sim 1000 \km \s^{-1}$, Becker {\it et al.} 1998);
and RX J0925.7-4758 ($5200 \km \s^{-1}$, Motch 1999).
The mass transfer rates in these systems is not known exactly, but is
probably above $\sim 10^{-8} M_\odot \yr^{-1}$ in all three sources.
We therefore adopt this value of $\dot M$ as the threshold for the
onset of CFWs in the case of white dwarf accretors.
Finally, we point out that the supersoft X-ray sources associated with
symbiotic novae are probably very closely related to the types
of systems we are considering in this work, especially during the
second phase of evolution when the secondary is an AGB star transferring
mass to its white dwarf companion (whose progenitor was the primary star).

  One of the processes that is likely to occur in many of the
systems studied in this work is a nova-like outburst of an accreting WD
companion, or a mass accretion instability of the accretion disk,
both onto WD and main sequence companions.
 Outbursts are observed in symbiotic systems.
Mikolajewska {\it et al.} (1999), for example, analyze the
symbiotic binary system RX Puppis, and argue for an orbital
separation of $\gtrsim 50 \AU$ between a Mira star and a
$\sim 0.8 M_\odot$ WD companion which undewent a nova-like
eruption during the last 30 years.
 Such an outburst can last up to 100 years  (Allen \& Wright 1988).
 Such an outburst will eject material at high velocities,
which can entrain nebular material and accelerate it to
several$~\times~ 100 \km \s^{-1}$.
 The high velocity material will be ejected by the WD preferentially along
the symemtry axis, or even if not, it will be
bound by the circumbinary dense material to expand in a cylindrical
or conical shaped flow along the symmetry axis.
 When observed several hundred years or more after the ``explosion'',
the fast material will show an approximate linear relation between the
distance from the central star and the velocity. 
 We suggest this explanation for this type of ``Hubble'' law flow observed
along the symmetry axis of several proto-PNs and PNs
(e.g., in MyCn18; Redman {\it et al.} 2000).

% ======================================================================
\subsection{Bending a Narrrow CFW}
% ======================================================================

 In order to analytically solve the equations describing the bending
process of the CFW, we take binary systems with large orbital separations,
though our basic results apply to closer systems as well.
  We consider a jet of a small opening angle $\alpha \ll \pi/2$, so that
the jet is well collimated (unlike a wide CFW), and all the material
located within a thin slice perpendicular to the jet's expansion
direction, will be bent simultaneously (unlike a wide CFW where
the side facing the slow wind bends first).
 The results give a good indication as to how the process
will work with wider jets as well.

 Let $h$ be the distance along the jet axis and perpendicular to the
orbital plane, and $v_f$ the het's velocity.
 We examine the region $ h \ll a$, so that we simplify the equations.
 The force per unit length on the jet at a distance $h$ from the mass
accreting star (secondary) is $F_{\rm ram} = P_{\rm ram} 2 h ~\tan \alpha$.
 Here $P_{\rm ram}= \rho v_s^2$ is the ram pressure of the slow wind
in the direction of the line joining the two stars, at the location of
the secondary, and $\rho_s$ the density of the slow wind.
 In the present simple calculation we can take the slow wind
to be spherical.
 Substituting for the slow wind's density we find
\begin{eqnarray}
F_{\rm ram} = \frac {\dot M_1 v_s}{4 \pi a^2}
2 h \tan \alpha.
\end{eqnarray}
 The distance that a segment of the jet progresses from the secondary
after a time $t$ is given by $h=v_f t$.
 The mass per unit length along the jet is
$m_j = \dot M_f /(2 v_f)$, where we divided by 2 since there are two jets,
one on each side of the orbital plane.
 Using the above expression, we find for the equation of motion
perpendicular to the jet (again, for $h \ll a$)
\begin{eqnarray}
\frac {dv_p}{dt}
=
\frac {F_{\rm ram}}{m_j}
=
\frac {\dot M_1 v_s}{\dot M_f v_f}
\frac {\tan \alpha}{\pi}
\frac {v_f^3}{a^2} t,
\end{eqnarray}
where $v_p$ is the velocity of the jet perpendicular to its initial
direction.
 Solving this equation with the initial condition $v_p(t=0)=0$,
and substituting $t=h/v_f$ and $\dot M_f = f \dot M_2$ we obtain
\begin{eqnarray}
\frac {v_p}{v_f} = 0.5
\left( \frac { \dot M_1 v_s } { \dot M_2 v_f } \right)
\left( \frac { \tan \alpha}{\pi \delta} \right)
\left( \frac { h^2}{a^2} \right),
\end{eqnarray}
where $\delta$ is the fraction of the mass accreted by the companion
which is blown into the jet.
 We now substitute the accretion rate from equation (3),
with $v_r \simeq v_s$, and use the explicit expression for the
accretion radius $R_a$ in calculating $\dot M_2$ (eq. 3).
This gives
\begin{eqnarray}
\frac {v_p}{v_f} = 0.03
\left( \frac {v_s} { 0.01 v_f } \right)
\left( \frac {v_s} { 15 \km ~s^{-1} } \right)^4
\left( \frac {M_2}{0.6 M_\odot} \right)^{-2}
\left( \frac { \tan \alpha}{f} \right)
\left( \frac { h}{10 \AU} \right)^2.
\end{eqnarray}
  Arranging it differently gives
\begin{eqnarray}
h \simeq   20
\left( \frac {v_p}{v_f} \right)^{1/2}
\left( \frac {v_s} { 0.01 v_f } \right)^{-1/2}
\left( \frac {v_s} { 15 \km ~s^{-1} } \right)^{-2}
\left( \frac {M_2}{0.6 M_\odot} \right)
\left( \frac { \tan \alpha}{10 f} \right)^{-1/2}
\AU.
\end{eqnarray}
 For orbital separations of $a \lesssim 10 \AU$, there will be high
mass loss rate in the equatorial plane due to tidal effects and
orbital motion, and less mass loss rate
per unit solid angle above the plane.
 This means a higher accretion rate and so a stronger CFW,
 whereas the wind bending the CFW is weaker.
 Therefore, the bending is not efficient, since when $h \gtrsim a$
the angle of the slow wind hitting the jet is high, hence reducing
$F_{\rm ram}$.
 The resulting possible flow structures are depicted in Figures 1 and 2,
and were discussed in the previous section.

% ======================================================================
\subsection{Elliptical PN Progenitors with CFW}
% ======================================================================

 If the orbital separation is $a \gtrsim 20 \AU$, the exact value depends on
the ratio $M_2/M_1$, but a CFW is still formed, then:
($i$)  Tidal interactions will be very weak, hence mass loss from the
       primary will initially deviate only slightly from sphericity;
($ii$) Accretion rates will be low,
       $\dot M_2 \lesssim 0.01 \vert \dot M_1 \vert$ (eq. 3), and
       hence the momentum of the CFW will be much less than that of the
       slow wind, and its energy comparable or less than that of the
       slow wind;
($iii$) From the equations (10)-(12) we see that the jet (or CFW) will
       be sharply bent at a distance of $h \lesssim a$ from its source
       (unless it is highly collimated).
 All these effects mean that the descendant PN will be elliptical,
i.e. no equatorial waist at all, but there will be a different
structure in, and near, the equatorial plane.
 The bent jet (or CFW) will not increase the momentum flux in the
equatorial plane, it rather will reduce it a little.
Its main effect will be to compress the matter in the equator,
as we discussed in the previous section.
  Other processes can form elliptical PNs with dense material
in the equatorial plane, e.g., a common envelope.
 However, the mass loss as the secondary enters the common envelope
 is likely to disrupt the elliptical structure in the polar direction.

 In the catalog of Manchado {\it et al.} (1996), containing 243 PNs,
we could find 3 PNs having the expected structure discussed above.
 These are JnEr 1 (PNG 164.8+31.1), A 70 (PNG 038.1-25.4)
and to less extent  M3-52 (PNG 018.9+04.1).
 It is possible that the fast wind blown by the primary central star
during the PN phase, will eventually destroy the original structure.
 Overall, we estimate that $\sim 1 \%$ of all PNs will belong to the
class discussed here, namely elliptical PNs with a signature 
near the equatorial plane of a CFW blown by a companion to the AGB star.

% ======================================================================
\section{POPULATION SYNTHESIS}
% ======================================================================
% ======================================================================
\subsection {Overview}
% ======================================================================

	In the previous sections we have described a scenario wherein two
stars of intermediate mass can evolve in a wide binary to produce bipolar
PNs with jet-like features and narrow waists.
 A central feature of our model is that while either of the stars is on
the AGB it may develop a stellar wind that is sufficiently large so as
to create an accretion disk around the companion star
(either a main sequence or white dwarf) with a sufficiently high mass
accretion rate that a CFW develops.
In order for this scenario to develop, the parameters
of the primordial binary must fall within certain ranges of parameter
space.  The question then arises as to how probable are the conditions for
forming a CFW, i.e., for what fraction of all PNs can we expect a CFW to
form?  We have taken a first step toward answering this question by
carrying out a population synthesis and binary evolution study of
primordial binaries that might potentially form such interestingly shaped
PNs.

	In our population synthesis and evolution study, we utilize Monte
Carlo techniques, and follow the evolution of some $5 \times 10^4$
primordial binaries.
For each primordial binary, the mass of the primary is chosen from an
initial mass function (IMF), the mass of the secondary is picked according
to an assumed distribution of mass ratios for primordial binaries, the
orbital period is chosen from a distribution covering all plausible
periods, and the orbital eccentricity, $e$, is chosen from a uniform
distribution.   Once the parameters of the primordial binary have been
selected, the two stars are evolved simultaneously using relatively simple
prescriptions (described below).  We explicitly follow the wind mass loss
of both stars at every step in the evolution.  For this purpose we have
developed a wind mass loss prescription that depends on the mass and
evolutionary state of the star, and that is designed to reproduce
reasonably well the observed initial-final mass relation for single stars
evolving to white dwarfs.  We also take into account the evolution of the
binary system under the influence of stellar wind mass losses.

	At each step in the evolution, we compute the fraction of the
stellar wind of one star that will be captured via the Bondi-Hoyle
accretion process by its companion.  In the case of the evolution of the
primary star, the wind will be captured by its main sequence companion,
while in the second phase of the binary evolution, the wind lost by the
original secondary star will be captured by the white dwarf remnant of the
original primary star.
 In addition to the mass capture rate, we also estimate whether sufficient
angular momentum will be accreted to allow for the formation of an
accretion disk before the accreted matter falls on the companion.
 We believe that this is essential for the formation of a CFW.
Finally, if the accreted matter forms a disk, we
check whether the total rate of accretion exceeds a certain critical value
(discussed below) to form a CFW.  If all of these conditions are met, then
various parameters of the binary system are recorded, e.g., the mass loss
rate of the AGB star, the mass accretion rate of the companion, the stellar
masses, the total envelope mass lost by the AGB star during the CFW phase,
the binary orbital parameters, and so forth.

	Finally, as the binary system evolves, we check at each step on two
other possible stellar interactions: (i) Roche-lobe overflow, and whether
it is stable or unstable, and (ii) tidal synchronization and circularization,
and whether it leads to a spiral-in merger due to the Darwin instability.
The prescriptions for handling these interactions are discussed below.

% ======================================================================
\subsection {Specific Prescription and Algorithms}
% ======================================================================

  The properties of the primordial binary systems are chosen via
Monte Carlo techniques as follows.
The primary mass is picked from Eggleton's (1993;
see eq. 1 of Di Stefano, Rappaport, \& Smith 1994) Monte Carlo
representation of the Miller \& Scalo (1979) IMF,
\begin{eqnarray}
M(x) = 0.19 x [(1-x)^{3/4} +0.032(1-x)^{1/4} ]^{-1},
\end{eqnarray}
where $x$ is a uniformly distributed random number
 This distribution flattens out toward lower masses, in
contrast with the Salpeter IMF (1955).
We considered primary stars whose mass was in
the range of $0.8 < M_1 < 8 M_\odot$.  Next, the mass of the secondary,
$M_2$, is chosen from one of several probability distributions, $f(q$),
of mass ratio (see, e.g., Abt \& Levy 1976, 1978, 1985;
Abt, Gomez, \& Levy 1990;
Tout 1991; Duquennoy \& Mayor 1991), where $q \equiv M_2/M_1$.
For our ``standard model'' we use $f(q) = Cq^{1/4}$, where $C$ is
a normalization constant.
This distribution has the property that the mass of the secondary is
definitely correlated with the mass of the primary, but is not strongly
peaked toward $q = 1$.
 Secondary masses down to $0.08 M_\odot$ are considered.
  To choose an initial orbital period, a distribution that is uniform
in log(P) over the period range $1$ days to $10^6$ years is used.
After the masses and orbital period are chosen, the orbital separation
is calculated from Kepler's law.
Finally, the orbital eccentricity is chosen from a uniform distribution
between 0 and 1.
 In a forthcoming paper we will further explore the sensitivity of
the results to different values of these and other parameters.
 We have already done some exploration of parameter space.

	The algorithm for evolving each of the two stars in the system is a
modified version of one used previously to evolve intermediate mass stars
(see, e.g., Joss, Rappaport, \& Lewis 1986; Harpaz, Rappaport, \& Soker 1997;
Rappaport \& Joss 1997).   Both the radius and luminosity are taken to be
simple functions of the core mass and total mass of the star.  These are
given by equations (5) and (6) of Harpaz et al. (1997), which are slightly
modified versions of a similar set given by Eggleton (1992).
 Since the growth of the core is
proportional to the luminosity, these expressions form a closed set of
equations for evolving the star.
 To the formulation we have used in the past, we have added a new
prescription for computing the mass of the stellar core
that is used in equations (5) and (6) of Harpaz {\it et al.} (1997):
\begin{eqnarray}
 m_{\rm core} = 0.12 M^{1.35}(Y-Y_0)/(1-Y_0) \qquad {\rm  for}
\quad Y < 1.
\end{eqnarray}
 The first part of this expression ($0.12  M^{1.35}$) is from
Terman, Taam \& Savage (1998; and references therein),
where, $M$ is the total stellar mass at the time when $m_{\rm core}$
is computed, $Y$ is the He mass fraction in the core,
and $Y_0$ is the initial He mass fraction.
 After $Y$ attains a value of unity, the core mass is taken to be that
when Y reaches unity plus any mass that is added to it through
nuclear burning.
We note, however, that evolutionary calculations based on the
above formalism do not incorporate special events during the stellar
evolution, such as the helium flash in the core at the tip of the first
red giant branch, the evolution along the horizontal branch, or
helium shell flashes (thermal pulses) during the AGB evolution.
They also do not yield a highly accurate
prescription of the stellar properties for the more massive stars we
consider (e.g., 3 to $8 M_\odot$).
  However, they do yield an adequate overall representation of the
evolution of the stellar core, radius, and luminosity.

	The wind mass loss rate from each star is computed for each time
step in the evolution.
 We have devised the following semi-empirical formula for the wind loss rate:
\begin{eqnarray}
\dot  M_{\rm wind} = f_R f_{sw} f_i
= 4 \times 10^{-13} ~L ~R ~M^{-1}
f_{sw}(R,M)f_i(M_0) ~M_\odot \yr^{-1},
\end{eqnarray}
where $f_{sw}$ is given by
\begin{eqnarray}
 f_{sw} = 1 + \exp (18 - 5000M/R),
\end{eqnarray}
and $f_i$ is given by
\begin{eqnarray}
f_i=
\exp(-4.45+3.308M_0 -0.8798M_0^2+0.09862M_0^3-0.004030M_0^4),
\end{eqnarray}
where in all these expressions the mass, luminosity, and radius are in
solar units.
 The first of these expressions $f_R$, is the usual Reimers' wind loss
formula (Reimers 1975), while the second term represents the enhancement
of the wind on the AGB during the ``superwind'' phase.
 The form of the second term is taken from the works of
Bedijn (1988), Bowen (1988), and Bowen \& Wilson (1991).
 The third term $f_i$, reduces the mass loss rate for low mass stars and
increases it for massive stars.
 It represents our ignorance of some processes that increase mass
loss rates for initially more massive stars, e.g., they tend to mix more
heavy elements into the envelopes by dredge-up, which makes dust formation
more efficient, and hence increases the mass loss rate.
 We adjusted some of the free parameters in $f_{sw}$ and in $f_i$ so as to
best reproduce the initial-final mass relation for the production of
white dwarfs in single stars (Weidemann 1993).
 The influence of the exact choice of values for these parameters on
the production of bipolar PNs will be studied in a separate paper.
  If $\dot M_{\rm wind}$ exceeds $3 \times 10^{-5} M_\odot \yr^{-1}$,
we fix $\dot M_{\rm wind}$ at this value.
 We also checked the case where the mass loss rate was limited to
the maximum mass loss rate possible via momentum transfer from radiation
$\dot M_{\rm max} = L /(c v_s)$, where $c$ is the speed of light and
$v_s$ the terminal wind velocity. We found no significant differences
in our results.

    As the stars in the binary evolve and lose mass through their
stellar winds, the orbit must also evolve.
  If the stellar wind is emitted isotropically in the rest frame
of the giant, and if the outflow velocity is much larger than the
orbital velocity, then the orbital eccentricity will remain constant
and the semimajor axis will grow as $da/a = \vert dm \vert /M_T$,
where $\vert dm \vert$ is the mass lost in the stellar wind and
$M_T$ is the total mass of the binary.
 In the systems we are considering, the wind speed during the AGB phase
will generally be comparable with the orbital speed.
In this case there can be a complicated interaction of the wind with the
orbit on its passage out of the system.
 To our knowledge, this dynamical problem has not yet been solved
analytically.
 Numerical simulations exist only for circular orbits and are
computed for only a limited range of parameters (MM98, MM99).
  We therefore use the following expressions for orbital evolution
which contain two adjustable parameters, $\alpha$ and $f_{ea}$:
\begin{eqnarray}
(da/a) = \vert dm \vert [M_1+2M_2(1-\alpha)][(1+f_{ea})M_1 M_T]^{-1},
\end{eqnarray}
and
\begin{eqnarray}
de=-(da/a)(1-e^2)e^{-1} f_{ea},
\end{eqnarray}
where $\alpha$ is the specific angular momentum carried away by the wind
material in units of the specific angular momentum of the wind losing star,
and $f_{ea}$ is a parameter which dictates how much the eccentricity changes
compared with the fractional change in semimajor axis;
$M_1$ is the mass of the mass-losing star and $M_2$ is that of the
companion.
 We note that these expressions do conserve overall angular momentum.
For our standard model, we take $\alpha = 1$ and $f_{ea} = 0$,
which reproduces the case of wind veloicty much larger than the orbital
velocity.
We have also run the code for a range of other values of these parameters
(i.e., $0.5 < \alpha < 3$; $0 <f_{ea} < 1$) in order to estimate the
sensitivity to these parameters.
 In a recent study, Hachisu, Kato, \& Nomoto (1999) conclude that
when the AGB wind speed is comparable to, or slower than, the orbital speed
the wind can carry away a large specific angular momentum.  This effect can
yield values of alpha that are considerably larger than the ones we have
tested, and will tend to cause many of the closer orbits to decay
dramatically rather than expand as with our standard model.  A preliminary
test of this somewhat extreme prescription for angular momentum loss
indicates that our results would be affected quantitatively, but not
qualitatively.  We leave a detailed exploration of this and other
prescriptions for angular momentum carried away by the AGB wind to a future
work.

  For each system we check if a strong tidal interaction takes place on
the RGB, i.e., if the synchronization time is shorter than the
evolutionary time.
We use the equilibrium tidal interaction (Zahn 1977; 1989;
Verbunt \& Phinney 1995) with time scales from Soker (1998a),
and eccentricity dependence from Hut (1982).
Neglecting the weak dependence on some variables, we take the
condition for a strong tidal interaction to be:
\begin{eqnarray}
R_{RGB} > 0.1 a_0 q^{-1/3} [f_s(e^2)]^{-1/6},
\end{eqnarray}
where $a_0$ is the initial orbital separation,
$R_{RGB}$ is the maximum radius the star attains on the RGB,
$q = M_2/M_{RGB}$, and
$f_s(e^2) \simeq [1+(15/2)e^2 + (45/8)e^4 + (5/16)e^6]/ (1-e^2)^6$.
The value of $R_{RGB}$ is taken from the results of Iben \& Tutukov (1985;
fig. 31), which we approximate by
$\log (R_{RGB}/R_\odot) = A (M/M_\odot) + B$, with different values of
$A$ and $B$ in 6 mass intervals.
 The boundaries of the 6 mass intervals between $0.8$ and $8 M_\odot$
are $M(M_\odot)=(1.5,2.1,2.2,2.3,5)$,
and the values of $A$ and $B$ in these intervals are
$A=(0,-0.2,-1.8,-9,0.267,0.16)$, and
$B=(2.3,2.6,5.96,21.8,0.486,1.02)$, respectively.
  We note the sharp change in behavior near $M=2.3 ~M_\odot$.

 As the two stars evolve in the binary the conditions for Roche-lobe
overflow (RLOF)  for both of the stars is checked.
Since the binary orbit is, in general eccentric, we take as a measure
of the size of the critical potential lobe for the onset of mass
transfer the quantity $ a (1 - e) f_{\rm Egg}(q)$,
i.e., the distance of closest approach at periastron times the
dimensionless function of mass ratio given by Eggleton (1983)
\begin{eqnarray}
f_{\rm Egg}(q) = 0.49 q^{2/3} [0.6 q^{2/3} + \ln (1+q^{1/3}]^{-1},
\end{eqnarray}
where $q \equiv M_1/M_2$ and $M_1$ is the star whose critical potential
is being evaluated.
 If either of the two stars overflows its critical potential lobe, computed
from the above expression, we test for the stability of the subsequent mass
transfer.
We then stop the evolution and record the system as having entered either a
stable or an unstable RLOF.
 The simple criterion for stability that we use is $q < 1$ at the epoch
of RLOF.
 We note though, that before the system enters the RLOF phase, the
system has a strong tidal interaction, that may substantially enhance
the mass loss rate (Tout \& Eggleton 1988).
 This enhanced mass loss rate will both increase the
orbital separation and reduce the primary star's mass, both of which make
the system more stable to a subsequent RLOF.
 To account for this effect, we also check the number of systems that
go through RLOF and have values of $ q \lesssim < 1.4$.
 As we discussed later, the stable RLOF systems, and some of the unstable
ones, are likely to form very narrow waist bipolar PNs.

     When the star reaches the AGB we check if tidal interactions can
bring the system into circularization and synchronization, and if positive,
we also check if the conditions for a Darwin instability are met.
 The Darwin instability will set in if two conditions are met:
 (a) the secondary cannot bring the primary envelope into corotation,
and
 (b) the spiraling-in time is shorter than the evolution time $\tau_{ev}$.
 If both conditions occur, a tidal catastrophe ensues.
 The condition for tidal catastrophe to develop
from synchronized orbital motion is (Darwin 1879) $I_{\rm env} > I_o/3$,
where $I_{\rm env} = k M_{\rm env} R^2$ is the envelope's moment of
inertia with $k \sim 0.2$ for giants, $M_{\rm env}$ is the envelope
mass, and $I_o = M_2 a^2$ is the orbital moment of inertia due largely to
the secondary. For $M_2 \ll M_1$, which is implied by condition (a), the
spiraling-in time is shorter than the evolutionary time along the upper AGB
if (Soker 1996)
\begin{eqnarray}
\frac {a}{R} \lesssim  5
\left( {{M_{\rm {env}}} \over {0.5M_1}} \right)^ {1/8}
\left( {{M_{\rm {env}}} \over {0.5M_\odot}} \right)^ {-1/24}
\left( {{M_2} \over {0.1M_1}} \right)^ {1/8} [f_c(e^2)]^{1/8},
\end{eqnarray}
where
$f_c(e^2) \simeq [1+(31/2)e^2+(255/8)e^4+(185/16)e^6)]/(1-e^2)^{15/2}$
(Hut 1982).
  In deriving equation (22) we use the equilibrium tide mechanism for
convective envelopes (Zahn 1977; 1989), and neglect the weak dependence on
some of the physical variables (e.g., stellar luminosity).
 The orbit is circularized on a timescale similar to the spiraling-in time.
 Therefore, if condition (b) above is met, the orbit will become circular.

     If both these conditions are met, then the system is registered as a
PN that went through a common envelope phase, and in most cases will form
an elliptical PN.  A bipolar PN might be formed in the following situation.
If (b) is met but not (a) the system comes into corotation. We continue to
check condition (a) and for Roche-lobe overflow.  If, as the primary
envelope expands, the companion enters a common envelope, either via tidal
catastrophe or Roche-lobe overflow, the system can still have formed
a bipolar PN before it went through a common envelope phase.
 In that case the final separation of the two post common envelope
stars (after the envelope is expelled) is relatively
large, as in the central binary system in the bipolar PN NGC 2346.  In this
case, even if a very-narrow waist is formed initially, it will not survive
after the common envelope phase.

	If the system does not enter a common envelope via either
Roche-lobe overflow or the Darwin instability, then we check if a CFW
forms.  The first condition for a CFW to form is given by equation (2),
where the specific angular momentum of the matter accreted by the companion
must be sufficient to form an accretion disk.
 We further require, by using equation (3), that the accretion rate
should be $> 10^{-8} M_\odot \yr^{-1}$, or
$> 10^{-7} M_\odot \yr^{-1}$, depending on whether the accreting
star is a white dwarf or main sequence star, respectively, in order for
the CFW to be strong enough, at least during part of the evolution.
  These limits on the accretion rates are based on supersoft X-ray sources
(for WDs; see $\S 3.1$), and on YSO, which accrete $\sim 1 M_\odot$
in $\sim 10^{7} \yrs$ (for main sequance accretors). 
  If these conditions are met, then the system is taken to have
formed a bipolar PN with a very narrow waist, or an elliptical PN with
equatorial prominent structure ($\S 3.3$).
 In the present calculations we did not consider the effect of the
accreted mass on the evolution of the orbital separation, and we did not
add the accreted mass to the secondary when we followed its subsequent
evolution.  These efects are important in only a small number of systems,
where the accretion rates are high. In these cases (of high accretion
rates) some other effects which are also potentially important are not
treated in the present work (e.g., gravitational focusing; MM99). "

% ======================================================================
\subsection {Probability of Forming Planetary Nebulae in Binary Systems}
% ======================================================================

    There are two additional questions to be answered before carrying out
the population synthesis calculations and making comparisons with the
observations:
(1) What is the fraction of all stars that are born in
binary (or triple) systems, and
(2) what is the probability that a star of initial mass $M_0$ forms a PN?
  The common answer to (1) is that $\sim 50 \%$ of systems are
binary, or triple, and $\sim 50 \%$ are single stars
(that may still have brown dwarf or planet companions).
However, other assumptions are also made.
Both  Han {\it et al.} (1995) and Yungelson {\it et al.} (1993) preferred
to consider that all stars are born in binary systems.
 In the present paper we analyze our results under the assumption
that the type of binary systems we are simulating (i.e.,
the primary masses are in the range of $0.8$ to $8 M_\odot$)
compose about half of all systems.
 That is, for each binary system, there is a single star having the same
properties as the primary star of the binary system.

     Both  Han {\it et al.} (1995) and Yungelson {\it et al.} (1993)
assume that all stars that are not disturbed by a stellar companion form
a PN.
 As we see now, this is not the case.
  The arguments that not all stars form PNs are summarized in detail in
a recent paper by Allen, Carigi, \& Peimbert (1998).
 They also argue in favor of giving the same weight to each PN, observed
or simulated, when comparing theory with observation, despite the fact
that PNs with massive central stars evolve faster.
 That is, the chance of observing a PN is the same for all types of PNs.
 From observations and a model they build, they find the probability
for a star with a given initial mass $M_0$ to form a PN
(their table 4).
 Their function has large uncertainties, and we simply
approximate it by the function
\begin{eqnarray}%
P_{\rm PN}=0.76 (M_0/M_\odot)-0.53  
\end{eqnarray}
where for $M_0<0.83 ~M_\odot$ the probability is set to zero, while it is
set to unity for $M_0>2 ~M_\odot$.
 The reason low-mass stars do not form PNs is that they have a high mass
loss rate on the RGB and early AGB.
 This may by caused by planets or brown dwarfs that spin up the star
as it evolves along the RGB (Soker 1998c; Siess \& Livio 1999),
or an interaction with a close companion, as are the cases here for
some systems that do not survive the RGB phase by the condition of
equation (20) above.
 We assume that stars of $M_0<1.6 M_\odot$ which have a strong RGB
interaction, do not form PNs.

 Because of the uncertainties regarding the exact answers to the two
questions posed above, we estimate the number of expected PNs in
the following simple way.
  We take $\beta_b \simeq 0.5-1$ to be the fraction of stars that
are born with a stellar companion, and therefore a fraction $1-\beta_b$
are born as single stars.
 We take the number of single stars that form PNs from the
function found by Allen {\it et al.} 1998 (eq. 23 above), as follows:
 We integrate equation (23) times the IMF over the mass of single
PN progenitors (in practice we used the population synthesis code
which is described in the next section) and find that $\sim 65 \%$ of
single stars with initial mass in the range $0.8 M_\odot < M_0 < 8M_\odot$
form PNs.
 To find the total number of PNs formed (for each initial star)
we multiply this fraction (of $0.65$) by $1- \beta_b$, and add the
fraction of PNs formed from the binary star systems we simulate in
the present work multiplied by $\beta_b$.

 As discussed above, binary systems that have a strong tidal interaction
during the RGB phase of the primary star are not evolved further
in the present version of our population synthesis code.
 These systems amount to $44 \%$ of the total binary systems we start with.
Somewhat arbitrarily, we assume that systems of this type
(that have a strong interaction on the RGB) for which the primary mass is
$M_1>1.6 M_\odot$, do eventually form PNs (e.g., stars merge via a
common envelope to form a more massive star which evolves as
a single star).
 The mass limit is taken arbitrarily to be two thirds of the mass
difference between $0.8 ~M_\odot$ (the lower limit for forming a PN)
and $2 ~M_\odot$
(a mass above which all stars form PNs according to Allen {\it et al.}).
The fraction of binary systems which have a strong tidal interaction on
the RGB and which we still consider will form a PN according to the
condition above, is $\sim 17 \%$ of all binary systems (see next section).
 From the systems we do evolve (those that did not have a strong interaction
on the RGB; see next section), we find that $\sim 33\%$ out of the total
number of initial binary systems
(including those with strong RGB interaction and those with very
large orbital separations which we do not folow here)
do form PNs, i.e., for each initial binary system we form $\sim 0.33$ PNs.

  Some low mass stars, $M_1<1.6 M_\odot$, with a strong interaction
on the RGB can also form PNs, e.g., a merger of the two stars will
form a more massive envelope from the destructed low mass star,
that may form a PN.
 Therefore the fraction can be larger than the $17 \%$ mentioned above.
 On the other hand, some wider systems with low mass stars which we do
evolve to the PN phase, may not form a PN.
 For example, a binary system with orbital separation of $10 \AU$
or more, can still possess a massive planet near its primary.
If the primary is a low-mass star, the planet may cause the
primary to lose all its envelope on the RGB.
We do not take account of such processes in our present simulations
of binary systems, and just assume that wide binaries do form PNs.
 In this way we overestimate the number of PNs formed in our simulation.
We assume that these two effects more or less cancel each other.

 Overall, we estimate that for each binary system there will be
$\sim (0.17+0.33)\sim 0.5$ PNs.
 Therefore, the expected total number of PNs for each
{\it binary system} is
\begin{eqnarray}%
F_{\rm PN} \simeq [0.65 (1-\beta_b) + 0.5 \beta_b](\beta_b)^{-1}
=(0.65-0.15\beta_b)/\beta_b.
\end{eqnarray}
 For $\beta_b=0.5$ we find that for every binary system
$\sim 1.15$ PNs are formed.
 Note that the number of PNs per each binary system can be larger than one
since we count here PNs which are formed from single stars,
as well as PNs which are formed from the initial primary, and in
some cases the initial secondary, of the binary systems (see next section).
 For $\beta_b=60\%$, i.e., $60 \%$ of all systems are binary systems
(Duquennoy \& Mayor 1991), the expected number of PNs for each binary
system is $0.93$.
 Overall, we find that for each binary system we expect
$\sim 1 \pm 0.15$ PNs to be formed.
  This is what Yungelson {\it et al.} (1993) and Han {\it et al.} (1995)
use for their analysis, but here we arrive at the same conclusion
from very different considerations.

% ======================================================================
\section{POPULATION SYNTHESIS RESULTS}
% ======================================================================
% ======================================================================
\subsection{Overall Statistics}
% ======================================================================

Once the population synthesis code was completed we made a number of runs
with it to study the fraction of binary systems that would produce bipolar
planetary nebulae, and especially those with very narrow waists.
Typical exploratory runs were carried out with 10,000 primordial binaries,
while in a few cases longer runs of approximately 50,000 systems were made.
All of the results presented in this section are based on a run of 46,700
systems.  In this case the primordial binary systems had primary masses in
the range of 0.8 to 8 Msun, and secondary masses down to 0.08 Msun, while
all other system parameters were chosen according the prescriptions
specified in Section 4.2.
The longer runs require about 5 hours of computation time on a
modest workstation.  Such runs produce about several 100 to 1000 of each of
the types of bipolar and elliptical PNs that are discussed later in this
section, thus the statistical accuracy in the numbers presented is usually
better than 10 percent.

The model parameters used to derive the results in this section are based
on our ``standard model''.
In particular we used $\eta = 0.2$ in eq. (2),
$v_s = 15  \km ~s^{-1}$ in eq. (4), the IMF given by eq. (13), a mass ratio
distribution given by  $f(q) = q^{1/4}$ (see section 5.2), stellar wind
parameters given by eqs. (15-17), evolution of the orbital eccentricity and
semimajor axis given by eqs. (18 \& 19) with $\alpha = 1$ and $f_{ea} = 0$.
We have also made runs for different values of $\eta$, $\alpha$, and
$f_{ea}$, but do not report those here, other than to note that the results
are not highly sensitive to the choice of these parameters. In all the
results presented below, unless noted otherwise, the units of mass,
velocity, orbital separation and accretion rate are in $M_\odot$, $\km
\s^{-1}$, $\AU$, and $M_\odot \yr^{-1}$, respectively.

  We start by analyzing the different classes we are interested in.
 The main classes are summarized in Table 2, and are defined below.
 The meaning of the different columns in Table 2 are:
(1) The class name. We do not give a class name to the systems with strong
tidal interaction on the RGB since we do not evolve them
(see $\S 4.2$ eq. 20). (2) Indicates which star is in the AGB phase
(i.e. the progenitor of the PN); ``Primary'' indicates the initially more
massive star of the binary system, while the ``Secondary'' is
the initially less massive component of the binary system.
(3) The type of interaction (see below for more detail). For example,
``CFW; no Sync'' means that a CFW is generated, but no orbital
synchronization/circularization is achieved.
(4) The total number of systems that belong to this class, expressed as a
percentage of the total number of {\it binary systems} we start with.
(5-7) The fraction of systems that belong to this class,
as a percentage of the total number of {\it binary systems} we start with,
and which form bipolar PNs (column 5), bipolar PNs with very narrow waists
(6), and elliptical PN with equatorial prominent structure due to the
CFW (7; see $\S 3.3$).

The definitions of the different classes are:
{\bf (A)} The initial primary is the progenitor of the PN, and the accretor
     is the main-sequence secondary star. The system generates a CFW, but
     no synchronization/circularization is achieved.
{\bf (B)} Like class A, but the initial secondary is the AGB star, and the
      accretor is the WD remnant of the initial primary star.
{\bf (C)} Like A, but the system reaches synchronization/circularization.
{\bf (D)} Like B, but the system reaches synchronization/circularization.
  In these 4 classes, no RLOF occurs.
A caveat is in order here regarding the circularization condition.
Circularization is more difficult to achieve than synchronization,
as its time scale depends on the ratio of stellar radius to orbital
separation as $(R/a)^{-8}$ (eq. 22), while the synchronization timescale
goes as $(R/a)^{-6}$ (eq. 20).
 Here we take the more conservative approach of requiring circularization
for the formation of bipolar PNs, i.e., that the mass loss rate from the
equator
of the AGB star, or the binary system, be much higher than from the polar
directions.
 We base this on the work of de Medeiros \& Mayor (1995, and
references therein) who compare single and binary stars having the
same rotational velocities, and find that the X-ray luminosity of
binary systems which have achieved circularization is much higher.
  They conclude that circularization is necessary for enhanced X-ray coronal
activity in binary systems, but not a sufficient condition.
 Although they studied main-sequence stars, we know of no better condition to
use for the formation of highly non-spherical
mass loss from AGB stars.
 It is possible that in some systems synchronization and orbital motion
(see below) will be sufficient to cause highly non-spherical mass loss
which will lead to the formation a bipolar PN.
 We therefore may miss some bipolar PNs.
{\bf (E)} The system develops a CFW, and later undergoes RLOF.
{\bf (F)} Systems that enter into a RLOF before a CFW is formed.
       We also include in classes E and F a very small number of systems
       that enter a common envelope phase due to the Darwin instability.
We find that in all E and F systems the AGB star is the initial primary.

     Each of the last two classes is subdivided into three subclasses:
      (1) Stable RLOF, where we test different stability criteria.
      (2) Unstable RLOF, but the core is massive enough to form a PN,
          and the secondary is massive enough to influence the mass
          loss to form a bipolar PN.
           Somewhat arbitrarily, we set the limits for these masses
          in the present paper to be $M_{\rm core} > 0.4 M_\odot$
         and $M_2>0.3 M_\odot$.
         In a forthcoming paper we will study these systems in
         more detail, taking into account the envelope mass
         at the time of RLOF, etc.
      (3) The rest of the RLOF systems which will form no PN, or will
      merge to form
          elliptical rather than bipolar PNs, depending on the masses
          of the stars.

 Some of the systems that belong to class A and B may have a high
equatorial mass loss rate due to the orbital motion alone.
 This happens if the orbital motion is of the order as the wind
velocity.
 Without the CFW these may become elliptical PNs.
 However, as we propose in the present paper, the CFW will compress
this equatorial material, and a bipolar nebula will be formed.
 We therefore examined the orbital velocity of the mass-losing star
in the category A \& B systems at the end of the evolution.
 In no class B system does the orbital velocity exceed  $4 \km \s^{-1}$,
and therefore all of them will form elliptical PNs, according to
the present model, with an equatorial structure due to the CFW.
 They will form no bipolar PNs (columns 5 and 6 of Table 2).
  Of the $1.8 \%$ of class B systems, $0.7 \%$ have initial secondary mass 
of $M_{20} > 1.6 M_\odot$, and we take this fraction to be the
number of systems that will form PNs (column 7). 

  Out of the $3 \%$ of class A, in $0.7 \%$ the mass losing star has
an orbital velocity of $>6 \km \s^{-1}$, most of these, $0.5 \%$, have
$M_0>1.6M_\odot$.
 We therefore estimate that $0.5 \%$ of the $3 \%$ systems
of class A will form bipolar PNS with very narrow waists,
these are indicated in column 5 and 6 of Table 2.
 $1.5 \%$ out of the rest of class A systems have initial masses of
$M_{10} \gtrsim 1.6 M_\odot$, and are likely to form PNs.
 We therefore estimate that $1.5 \%$ (colum 7 of table 2) will form
elliptical PNs with a pronounced equatorial structure formed from
the CFW ($\S 3.3$). 

All (or most) systems of classes C and D are likely to
form bipolar PNs with very narrow waist.
 The numbers are given in columns 6 and 7 of Table 2.

 Of the $3.3 \%$ of systems that have RLOF (class E and F),
$0.42 \%$ ($1.4 \%$) are stable to RLOF under the condition
$q<1$ ($q<1.4$) at the onset of RLOF (where $q$ is the ratio of
the RGB star mass to the companion mass).
 Of these $3.3 \%$, $2.6 \%$ enter the RLOF stage with a core mass of
$>0.4M_\odot$ and a secondary mass of $M_2>0.3 M_\odot$.
 There are overlaps between these two groups.
 Many of these systems are likely to form bipolar PNs, a process we
do not study in the present paper.
 Based on these numbers and the masses of the progenitors (using
Allen {\it et al. } 1998), we estimate the fraction of
class E and F systems that do form bipolar PNs (see column 5 of Table 2),
and those who will also blow CFWs and form very-narrow waists bipolar PNs
(see column 6 of Table 2).
 These are crude estimates, since we did not follow these systems
after the onset of RLOF.
 Population synthesis of systems which evolve through a common envelope
phase and form PNs were done by
Yungelson {\it et al.} (1993) and Han {\it et al.} (1995).

  Other systems that may form bipolar PNs and which we do not evolve
are those which have a strong interaction while the primary star is
evolving up the RGB.
 In total, $44 \%$ of the binary systems belong to this class,
which is listed in the 7th row of Table 2.
 These systems will be studied in more detail in a forthcoming paper,
and we do not present details of their properties here.
 What is relevant to us is that of these systems,
$5.4 \%$ ($3.6 \%$) have a primary with an
initial mass of $M_{10}>1.6 M_\odot$, and an
initial orbital separation of $a_0>1 \AU$ ($a_0>2\AU$).
 Many of these are likely to form bipolar PNs.
Therefore, we estimate that $\sim 5 \%$ among these systems
do form bipolar PNs.
 Those with large separations and high mass loss rates at early stages,
may also blow a CFW at later stages.
 Since we did not evolve these systems, we crudely estimate the fraction of
these systems that will form very-narrow waists bipolar PNs to be
$\sim 1 \%$, (column 6 of Table 2).
 We note that there are PNs and proto-PNs with high density equatorial matter,
which have binary nuclei with orbital separations of $\sim 1 \AU$
(e.g., the  Red Rectangle; Jura, Turner \& Balm 1997 and references therein).

 We are now in a position to add all the numbers in Table 2, to estimate the
number of bipolar PNs according to the binary model (Morris 1987;
Soker 1998a), and the total of which have very-narrow waists according
to the scenario proposed in the present paper.
 We emphasize again that by binary model we refer to the model where
the stellar companion is outside the envelope during a substantial
fraction of the AGB phase of the mass-losing star (Soker 1998a).
 Common envelope evolution will lead in most cases to the formation
of elliptical PNs.
Thus we find that for every 100 binary systems $\sim 10$ bipolar PNs
are produced, with $\sim 1/2$ of them having very narrow waists
(last row of Table 2).
 Considering the total fraction of binary systems among the progenitors
of PNs (see $\S 4.3$), we find that our simulations imply that
$9-11 \%$ of all PNs are bipolar with $\sim 1/2$ of them
($4-5.5\%$ of all PNs) having very-narrow waists.

 Of the 43 bipolar PNs summarized in table 1, the number of PNs
which have very narrow waists is $13$, while for another 6 we cannot
tell. Therefore, the fraction of very narrow waists among the bipolar
PNs is $(16 \pm 3)/43=37 \pm 7 \%$.
 Since some very narrow waists might become wider at later stages,
we may miss some bipolar PNs which had very narrow waists at earlier
stages of their evolution.
 Overall, we estimate that about half of all bipolar PNs have, or had,
very-narrow waists.
 Since $\sim 11 \%$ of all PNs are bipolars (Corradi \& Schwarz 1995),
we find that $\sim 5.5 \%$ of all PNs are bipolar with very narrow
waists, with a large uncertainty, i.e., the number can be as low
as the observed fraction of $\sim 4 \% $, or much above $5.5 \%$,
if many very narrow waist PNs are destroyed at early stages of the
PN phase.
 We find the agreement between the results of the population synthesis
and the observations very encouraging for the binary model of
bipolar PNs, where the companion is outside the AGB envelope,
and for our proposed scenario for the formation of very-narrow waists
bipolar PNs.

 Most stars in classes A and B will form elliptical PNs with some
signature of the CFW near the equatorial plane of
the descendant PN ($\S 3.3$).
 Considering the large uncertainties, e.g., in the value of $\eta$
(see eq. 2), we estimate their number to be  $\sim 2 \%$ of all
binary progenitors (7th column of Table 2).
 This is compatible with the value estimated from observations at
the end of $\S 3.3$ ($\sim 1 \%$).
  We note also that some of the PNs here (colum 7 of Table 2),
may actually form bipolar PNs, if the CFW is strong and/or synchronization
rather than circularization is sufficient for enhanced equatorial mass loss
(see discussion above), increasing somewhat the expected number of
bipolar PNs. 

 Another success of the binary model for the formation of bipolar PNs,
where the companion is outside the AGB envelope during most of
the evolution, is the progenitors mass distribution.
 Our simulations show that the majority of progenitors of
bipolar PNs which we did evolve have high initial mass
($M_0\gtrsim 1.6 M_\odot$).
 To these we add the systems which undergo strong tidal interaction
on the RGB, and have a high-mass primary star and large orbital separation,
which we assume form bipolar PNs.
 These show that the binary model for the formation
of bipolar PNs can account for the observation that most
bipolar PNs come from high mass progenitors
(spectral type $\sim$F or earlier).
  However, our results also show that a small number of low-mass stars
may also form bipolar PNs.
 These may come from some class A systems which have a low mass primary,
but for which the orbital velocity of the mass losing star is large,
$v_1 \gtrsim 6 \km \s^{-1}$. 

% ======================================================================
\subsection{The Synchronized Systems with CFW}
% ======================================================================

	To better understand the types of systems we are modeling as the
progenitors of very narrow-waist bipolar PNs (VNW-BPNs), we show a series
of distributions of their parameters in Figures 3 through 7.
 These distributions are aimed at elucidating the properties of those
systems we classify in categories C and D in Table 2, i.e., those which
we believe will produce many of the very narrow waist PNs.
Recall that these are systems in which one (or eventually both) of
the stars in the binary accretes matter from its companion AGB star
wind and produces a strong CFW, which in turn pushes on the AGB wind
to produce a very narrow waist bipolar PN.
 All of the figures in this series have two panels which give the
distributions of parameters for the case where the primary is the AGB
star (Category C, panel a marked with ``Primary AGB''), and the
secondary is in the AGB phase (Category D, panel b marked ``Secondary AGB'').
  In the latter case, the accretion is onto the white dwarf
companion whose progenitor was the original primary in the binary system.

	The distributions of initial progenitor mass of the stars which
produce category C \& D systems are shown in Figure 3.  The darker histogram
in each case represents the original mass of the AGB star, while the
lighter histogram is the current mass of its companion star which produces
the CFW.
 Note in panel (a) that only primary stars with mass $\gtrsim 1.8
M_\odot$ produce very narrow waist bipolar PNs.
 This results from the fact that lower mass stars reach large radii
on the RGB and, through interactions with their companion, lose much
of their envelopes before ascending the AGB (see also Soker 1998a).
Note also that the companion masses in panel (a) can have rather modest
masses (e.g., down to $\sim 0.3 M_\odot$) and still produce a CFW
according to our model.
 From panel (b) we see that only secondaries with mass $> 1 M_\odot$
will produce VNW-BPNs, since lower mass stars will not have sufficient
time to evolve to the AGB or do not produce a sufficiently strong
stellar wind.
 Note that the lower mass companions in panel (b) are the white dwarf
remnants of the primaries in panel (a). 
The results shown in this figure help explain why the
progenitors of bipolar PNs have larger progenitor masses as inferred
from their z distribution in the Galactic disk
(e.g., Greig 1972; Corradi \& Schwarz 1995).

	The masses of the white dwarfs produced in our simulated VNW-BPN
systems are shown in Figure 4.  The msss distribution of white dwarf
remnants of the primary is similar to that of the secondary, except the
entire distribution of the latter is shifted to lower masses by
$\sim 0.1 M_\odot$.
  This can be understood from the fact that the secondary masses are
systematically smaller than those of the primaries, and hence the
respective production of white dwarfs will be somewhat different in the two
cases.

        The distributions of semimajor axes for the systems in Categories
C \& D are shown in Figure 5.  In each panel, the darker histogram
represents the initial semimajor axis while the lighter histogram indicates
the semimajor axes at the time of the PN formation.  Formation of Category
C \& D systems generally requires initial semimajor axes in the range of
$\sim 5-30 \AU$.
  At time of PN formation, these semimajor axes have increased to
$\sim 10-100 \AU$ in the case of the primary star reaching the AGB,
and $\sim 20-150 \AU$ for the case of the secondary ascending the AGB.
This orbital expansion is
due to the mass loss from the AGB stars (see equation 18).  Note, that
according to our proposed scenario, the Sirius A/B system, with an orbital
period of 50 years and a semimajor axis of $\sim 20 \AU$, is predicted
to produce a Category D PN with a very narrow waist in the future.

	The accretion rates onto the companion stars from the AGB winds of
the PN progenitors are shown in Figure 6. Accretion rates onto the
main-sequence companions (panel a) range from the minmum value that we
assume can produce a CFW ($\sim 10^{-7} M_\odot \yr^{-1}$) to a maximum of
$\sim 3 \times 10^{-5} M_\odot \yr^{-1}$.  The latter high values occur
when the companion is very near the AGB star so that it captures a
substantial fraction of its stellar wind.  The accretion rates onto the
companion white dwarfs (panel b) are typically lower due to the larger
orbital separations, the lower mass of the white dwarfs compared to typical
main sequence companions (hence the smaller Bondi-Hoyle capture radius),
and the fact that we impose a minimum value for the accretion rate of $\sim
10^{-8} M_\odot \yr^{-1}$ in order for a white dwarf to produce a CFW.

	Finally, in this series of distributions, we show in Figure 7 the
total mass lost by the AGB star during the CFW phase, i.e., the epoch
during which the very narrow waist bipolar PN is formed.  When the primary
is the AGB star, the total mass loss into what will become the PN ranges
all the way from a fraction of a solar mass to several solar masses.
However, we note that some of this material will be lost at a sufficiently
early epoch (e.g., $\sim 10^{5} \yr$ prior to the PN) that it will have
dissipated into the interstellar medium by the time the hot core of the
AGB star can fluoresce the nebula.
The mass lost by the secondary star's ascent of the AGB is
more tightly constrained to be $\sim 0.2 - 1.5 M_\odot$.

	We now present a series of Figures (8 through 12) which show
correlations between pairs of parameters for the systems in our study that
end up in Categories C \& D (the VNW-BPNs).  Figure 8 shows the final vs.
initial mass of the primary star (panel a) and secondary star (panel b).
Each dot in the plots represents one system at the end of the AGB phase.
Not surprisingly, these plots reproduce the initial-final mass curves of
Weidemann (1993); that is, we have adjusted our stellar wind prescription
(see equations 15-17) so as to ensure this result.

	The relation between the accreted mass and the accretion rate onto
the companion star is shown in Figure 9.  In the case of a main sequence
companion, we see that in at least $50 \%$ of the cases more than
$\sim 0.5 M_\odot$ has been accreted.
 This would explain why some of the bipolars have highly enhanced
abundances of N (see, e.g., Corradi \& Schwarz 1995; Pottasch 2000)
since, when these companions evolve to become AGB stars,
their envelopes will have been enriched by the mass transfer during
the previous epoch when the primary was an AGB star.
 By contrast, the larger orbital separations and smaller
masses of the white dwarf companions, lead to relatively small mass
accretions during the second phase of mass transfer (panel b).

	The variation of accretion rate with semimajor axis is shown in
Figure 10.  For the white dwarf accretors (panel b) there is a rather
simple dependence of $\dot M$ on semimajor axis which goes approximately
like $a^{-2}$.  This results from the fact that the terminal AGB stars are
nearly all emitting winds at a rate of $\sim 3 \times 10^{-5} M_\odot
\yr^{-1}$ in our simulations, and that the white dwarf masses are
$\sim 0.7 M_\odot$; therefore, the Bondi-Hoyle capture rate depends
essentially on the inverse square law assumed for the AGB winds.  By
contrast, the accretion rate-semimajor axis plot for main-sequence
accretors looks more like a "scatter diagram", due to the large range in
companion star masses (i.e., covering nearly two orders of magnitude).

	The relation between the initial masses of the secondary and
primary in Category C \& D systems is displayed in Figure 11.  By the
definition of ``primary'' and ``secondary'', all systems lie below a line
running at 45 degrees across the diagram.  However, these plots do
reemphasize two features seen in Figure 3.  Note that only systems where
the primary mass is $\gtrsim 1.6 M_\odot$ go on to become VNW-BPNs, and
that only those systems with secondaries with $\gtrsim 1 M_\odot$ go on to
produce a second phase of a VNW-BPN in the same binary system.  Other than
for these two restrictions, the remainder of the mass distribution simply
follows that of the IMF (see eq. 13).

	Finally, we show some effects of orbital eccentricity in Figure 12.
Note that for modest eccentricities ($\lesssim 0.5$) that the initial
semimajor axes must lie in a restricted range between $\sim 4$ AU and $\sim
15$ AU.
 By contrast, for larger eccentricities the initial semimajor axes
can be as large as about 50 AU, and there is a distinct lower limit
to $a_i$ which increases with increasing eccentricity.
 The lower limit results from the fact that the AGB star needs to fit
within its critical potential lobe at periastron
[proportional to $(1-e) a$],
 and the condition of no strong interaction on the RGB,
while the maximum values of the semimajor axis are determined by
the requirement for the specified minimum accretion rates that
are needed to produce the CFW.

% ===================================================================

% ======================================================================
\section{SUMMARY}
% ======================================================================

Morris (1987) suggested that a collimated wind blown by a companion to an
AGB star can influence the structure of the descendant PN and form a
bipolar PN (recall that Bipolar PNs are defined as PNs having two lobes
with an `equatorial' waist between them). This flow, which we have termed a
collimated fast wind (CFW), is the subject of the present paper.  Based on
observations, we estimate that $\sim 10\%$ of all PNs are bipolar,
and about half of these have very narrow waists.  It is these objects that
our scenario addresses. Our main claims, conclusions, and results are
summarized below.
\newline
(1) We propose that a CFW, if strong enough, will both compress the slow
wind near the equatorial plane and accelerate material along the symmetry
axis, and hence will lead to the formation of a bipolar PN with a very
narrow waist.  In our scenario, we further require that there be a strong
interaction, e.g., tidal interaction, between the AGB star and its
companion in order to have the AGB wind enhanced in the equatorial plane
 Another possibility for enhanced equatorial flow is a large orbital
velocity of the mass losing star (see text).
\newline
(2) If there is a CFW but no strong tidal interaction or fast orbital
motion of the mass losing star, then we suggest that
an elliptical PN will be formed.  If the companion remains outside the AGB
envelope, but still close enough to accrete mass, form a disk, and blow a
CFW, then a prominent structure should be formed in the equatorial plane of
elliptical PNs.  From observations, 3 PNs from a sample of 243,
we estimate that $\sim 1 \%$ of all PNs have such a structure.
 From the population synthesis study we find that $\sim 2 \%$ are likely
to belong to this class.
 Considering the large uncertainties due to the observations (only 3
objects) and our ignorance of the exact conditions for forming a CFW,
the two numbers are compatible
\newline
(3) We have quantified the conditions required for the formation of an
accretion disk around the companion to the AGB star, as well as those
necessary for the production of a CFW.  These include limits on the
specific angular momentum of the accreted matter, and the accretion rates
into the disk, as a function of the binary system parameters.
\newline
(4) Qualitative descriptions of the interactions of the CFW with the slow
wind from the companion AGB star are given (Figs. 1 and2 ), as well as
quantitative expressions for the bending of the CFW and the outflow
velocities of the accelerated shells of matter.
The geometry of the outflows in 3 dimensions
is likely to be even more complex than the picture we have sketched; a full
understanding of the dynamics and the effects of the wind interactions will
require future hydrodynamical simulations.
\newline
(5) We have devised a population synthesis code to follow up to 50,000
primordial binaries through the evolution of both stars, including wind
losses and orbital evolution.  The code was utilized to ascertain the
fractions of binary systems that would form PNs with the types of CFW/AGB
wind interactions that we have postulated.
\newline
(6) The population synthesis study shows that $\sim 10 \%$ of all PNs could
be shaped by a CFW to form a bipolar PN and that about one half of these
would be expected to have very narrow waists.  These fractions are in
accord with the values deduced by us from the observations.  In citing
these results, we have estimated that for each binary system in our
population synthesis study, the total number of PNs formed is $1 \pm 0.15$.
This includes both PNs formed from binary and single star systems.
\newline
(7) The majority of the systems that would form very narrow waist bipolar
PNs have strong tidal interactions or RLOF on the AGB, and descend from
progenitors having main sequence masses of $M_0>1.5 M_\odot$.  This later
point is in agreement with the observation that a very large fraction of
bipolar PNs are found near the Galactic plane and are presumed to
result from more massive stars.
\newline
(8) We estimate that a comparable number of additional bipolar PNs would be
formed from systems that we did not follow in the present work (a subject
of a forthcoming work).  These are systems that have strong interactions at
earlier stages, e.g., on the RGB.
\newline
(9) Our results also show that a small number of low-mass stars may also
form bipolar PNs.  Future more detailed calculations should follow these
systems.  We note that in the list of 43 PN presented in Table (1), M3-2 is
located at a relatively high galactic z of $0.7 \kpc$ (Corradi \& Schwarz
1995), which is typical for low-mass stars. IC 4406 is located at $0.46
\kpc$ from the galactic plane.  Both PNs have a bipolar structure, but a
very weak one, i.e., no very narrow waist and no large lobes.  We
hypothesize that these bipolar PNs may result from low-mass stars.
\newline
(10) We find from the population synthesis study that in at least half the
systems which form very narrow waist PNs that the secondary accretes at
least $0.5 M_\odot$ from the primary AGB star.  For those systems that go
on to produce a second PN phase when the secondary star evolves, we expect
such PNs to exhibit chemical enrichments of N (and some other elements).
 We therefore hypothesize that bipolar PNs with high metalicity are
descendant of stars that accreted mass from an initialy more massive
companion, and now have a WD companion to their central star.
\newline
(11) We also find that primordial binaries with semimajor axes in the range
of 5 - 30 AU are most likely to form bipolar PNs with very narrow waists;
these leave remnant binaries with semimajor axes of $\sim 10 - 100 AU$
(e.g., the Sirius A/B system).  If the binary goes on to produce another PN
when the secondary ascends the AGB, the final semimajor axes could range
from $\sim 20 - 200 AU$.  We note, however, that this conclusion depends
somewhat on the amount of specific angular momentum that the AGB wind
carries away from the binary system.
\newline
(12) Overall, we conclude that the population synthesis results are
compatible with the basic suggestion that bipolar PNs are formed in binary
systems where the companion to the AGB star spends a large fraction of the
AGB evolution outside the envelope. These results further support the
general scenario for the formation of asymmetrical PNs in binary systems.

%====================================================================
%====================================================================
\bigskip

{\bf ACKNOWLEDGMENTS:}
 We thank Amos Harpaz for a number of helpful discussions, and Goce
Zojcheski for his help with the initial phases of the calculations.  This
research was supported in part by grants from the 
Israel Science Foundation, the US-Israel Binational Science Foundation,
and NASA under its Atrophysics Theory Program: Grants NAG5-4057 and
NAG5-8368.
% ===================================================================

% ===================================================================

{\bf FIGURE CAPTIONS}

\noindent {\bf Figure 1: ask the authors for this figure}
A schematic drawing of the flow structure in the strong CFW case,
$\theta_d < \theta_c$.
 CFW stands for collimated fast wind blown by the compact secondary
at a velocity $v_f$.
 The figure shows the plane perpendicular to the equatorial plane, and
momentarily containing the two stars.
 The meaning of $L \sim 10 a$ is explained in the text, where $a$ is the
orbital separation.
 The thick line represent a shock wave.
We did not draw the shock waves of the slow wind as it hits the CFW,
and of bent CFW material.
\newline
(a) The inner region.
 The double-lines arrows represent the accretion flow onto the secondary,
where the long double-line arrow represent the accretion column behind the
secondary.
\newline
(b) An extended view, up to $ \sim 20 a$.
 The different regions are:
(A) The shocked CFW. No hot bubble is formed, as the hot post-shock
    material is leaking along the symmetry axis (region E).
(B) A compressed segment of a shell, formed from the interaction of
    the CFW with the slow wind in region C.
(C) Slow wind material ejected during the preceding orbital period.
(D) Low density region, filled with tenuous material streaming from the
    inner region of the shell F.
(E) Undisturbed (or only little disturbed) CFW flowing along the symmetry
    axis.
(F) The segment of a shell formed half a period earlier than the segment A
    i.e., a later stage of region B.  The segments F and B are connected
    by material outside the plane of the figure (a corkscrew structure).
(G) The high density region near the equatorial plane formed from the
    backflowing shocked CFW material, region H.
(H) Back flowing shocked CFW material, which compresses the slow wind
    gas toward the equatorial plane.
(I) The CFW entrains material from the edges of the dense shells, and
    accelerate it to high velocities.

\noindent {\bf Figure 2: ask the authors for this figure}
 Like Figure 1, but for the weak CFW case,
$\theta_d > \theta_c$.
 The different regions in (b) have the following meaning:
(A) The shocked CFW. A hot bubble is formed.
(B) A compressed segment of a shell, formed from the interaction of
    the CFW with the slow wind in region C.
(C) Slow wind material ejected during the preceding orbital period.
(D) Low density region, which is the shocked CFW material, i.e.,
    a later stage of region A. Regions A and D are connected
    with material outside the plane of the figure, in a corkscrew
    structure.
    It exerts a force on the outer shell F, on region E, on the
    freshly ejected slow wind in region H, and on the dense ring near
    the equator (region G).
    As it expands, and due to heat conduction to the shell F,
    this material cools.
(E) Undisturbed (or only little disturbed) slow wind material,
    along the symmetry axis.  On the edges, the material is compressed
    by the hot bubble (regions A and D).
(F) The segment of a shell formed half a period earlier than the segment A
    (a later stage of region B).
(G) The high density region near the equatorial plane formed from the
    pressure exert by the hot bubble (regions A and D).
(H) The slow wind material. It is expanding against the pressure of
    bubble D.
    Instabilities are expected in the interface between regions H and D.

\noindent {\bf Figure 3:}
Distributions of the initial mass of the AGB star and the mass of its
companion star at the end of the AGB phase for systems in Categories C and
D in our population synthesis.  These are the systems enumerated in Table 2
which have collimated fast winds (CFWs) and where the AGB star comes into
synchronous rotation with the orbit, i.e., our principal candidates for
forming very narrow waist PNs in binary systems.  The top panel refers to
such systems that formed when the original primary star in the binary
ascends the AGB and the companion is the original secondary (Category C),
while the lower panel is the same for the case where the secondary is
an AGB star and the companion is the white dwarf remnant of the
original primary (Category D).
 Note that the real number of the WD companions in each bin was
raduced by a factor of 3. 

\noindent {\bf Figure 4:}
Distribution of final core masses of the AGB stars that give rise to
systems in Categories C and D (see the caption of Figure 3 for a
description).

\noindent {\bf Figure 5:}
Distributions of initial and final semimajor axes for the systems in
Categories C and D (see the caption of Figure 3 for a description).  The
designation 'initial' refers to the start of the primordial binary, while
'final' refers to the end of the AGB phase of the primary (top panel) and
secondary (bottom panel).

\noindent {\bf Figure 6:}
Distribution of mass accretion rate onto the companion star from the
stellar wind of the terminal phase AGB star for systems in Categories C and
D (see the caption of Figure 3 for a description).  In the case where the
original primary is the AGB star, the accretion takes place onto its
main-sequence companion, while for the case where the secondary is the AGB
star the accretion is onto the white dwarf remnant of the original primary
star.

\noindent {\bf Figure 7:}
Distribution of total mass lost by the AGB star during the CFW phase for
systems in Categories C and D (see the caption of Figure 3 for a
description).  Note that the CFW originates from the accretion disk
surrounding the companion to the AGB star.

\noindent {\bf Figure 8:}
Final core mass of the AGB star vs. initial mass of that star for systems
in Categories C and D in our population synthesis.  These are the systems
enumerated in Table 2 which have collimated fast winds (CFWs) and where the
AGB star comes into synchronous rotation with the orbit, i.e., our
principal candidates for forming very narrow waist PNs in binary systems.
 Each dot represents one binary system.  The left panel refers to such
systems that formed when the original primary star in the binary ascends
the AGB (Category C), while the right panel is the same for the case where
the secondary is an AGB star (Category D).

\noindent {\bf Figure 9:}
Final accretion rate vs. total mass accreted by the companion of the AGB
star for systems in Categories C and D  (see the caption of Figure 8 for a
description).  In the case where the original primary is the AGB star the
accretor is its main sequence companion.  When the original secondary is
the AGB star the accretor is the white dwarf remnant of the original
primary.   If the accretion rate onto the white dwarf exceeds a
few$~ \times 10^{-7} M_\odot \yr^{-1}$,
the luminosity from nuclear burning may exceed
the Eddington limit  and much of the accreted matter would not be retained
in the system; in this case 'accreted mass' and 'accretion rate' refer only
to matter that enters the accretion disk.  In all cases, only a small
fraction of the accreted mass is assumed to be ejected in the CFW.

\noindent {\bf Figure 10:}
Final accretion rate vs. semimajor axis for systems in Categories C and D
(see the captions of Figures 8 and 9 for a description).

\noindent {\bf Figure 11:}
Initial mass of seconday  vs. initial mass of primary for systems in
Categories C and D  (see the caption of Figure 8 for a description).  Note
that only primary stars with mass $\gtrsim 1.6 M_\odot$ go on to form systems
with CFWs and narrow waists.

\noindent {\bf Figure 12:}
Orbital eccentricity  vs. initial semimajor axis for systems in Categories
C and D  (see the caption of Figure 8 for a description).

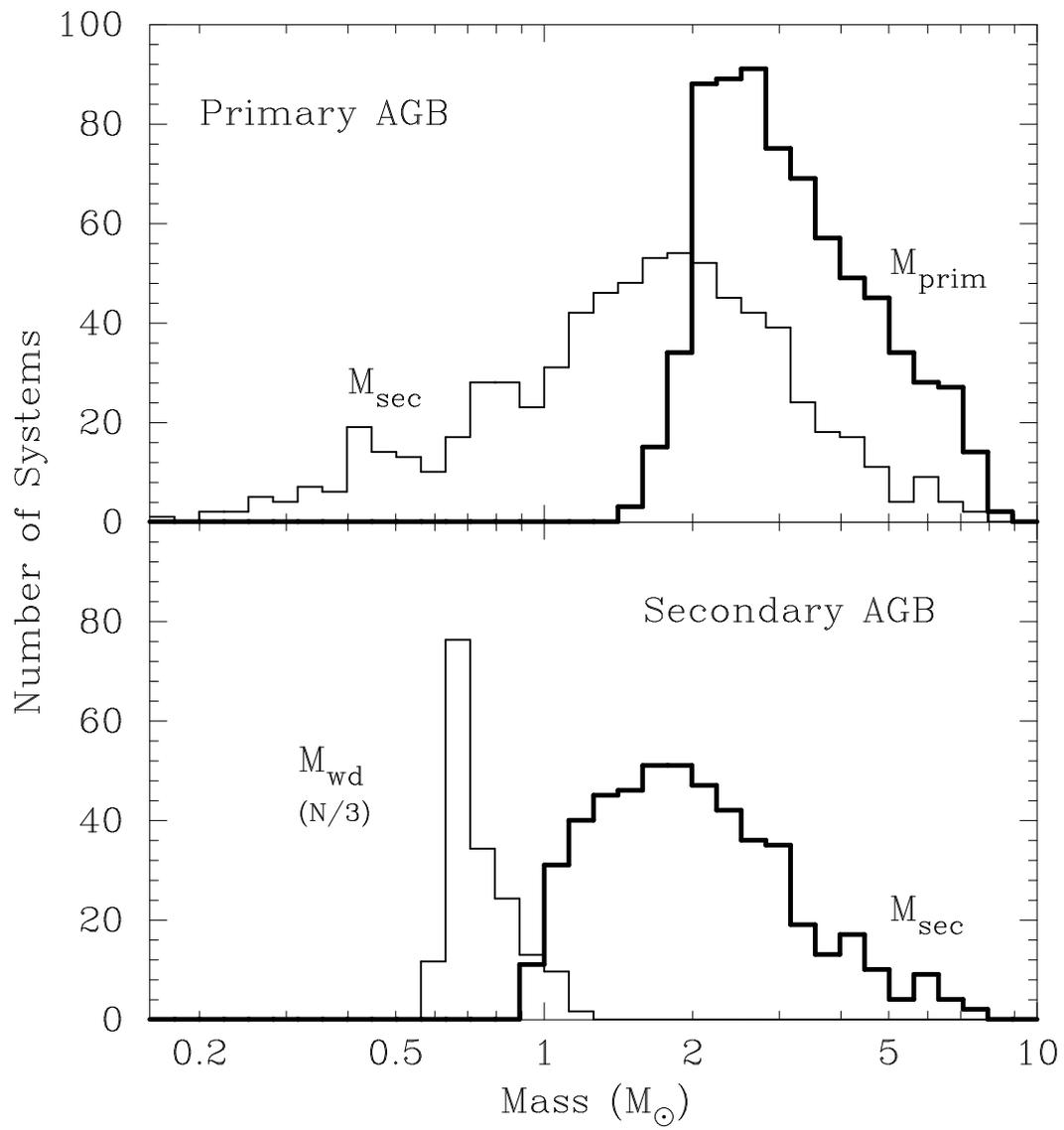
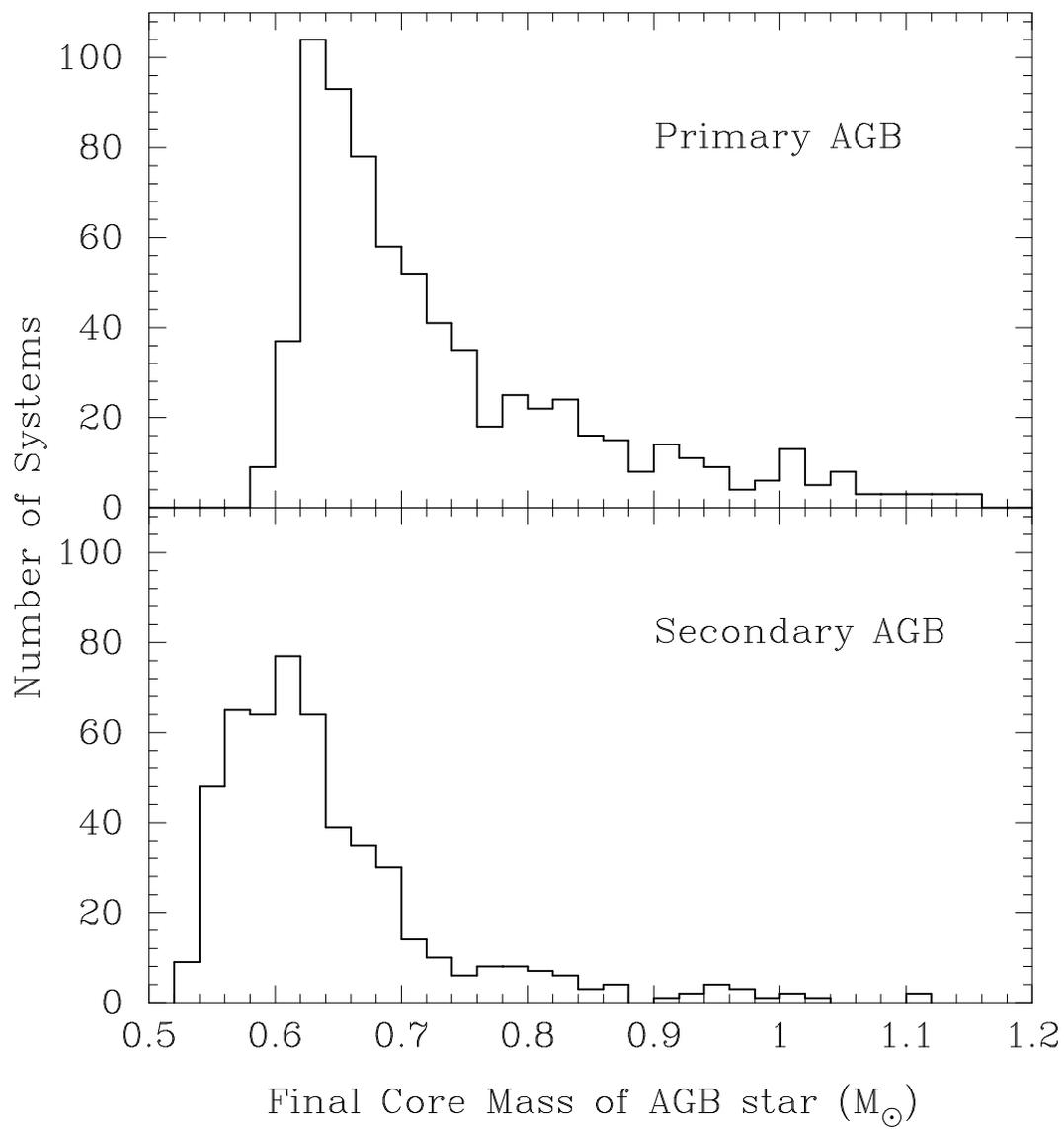
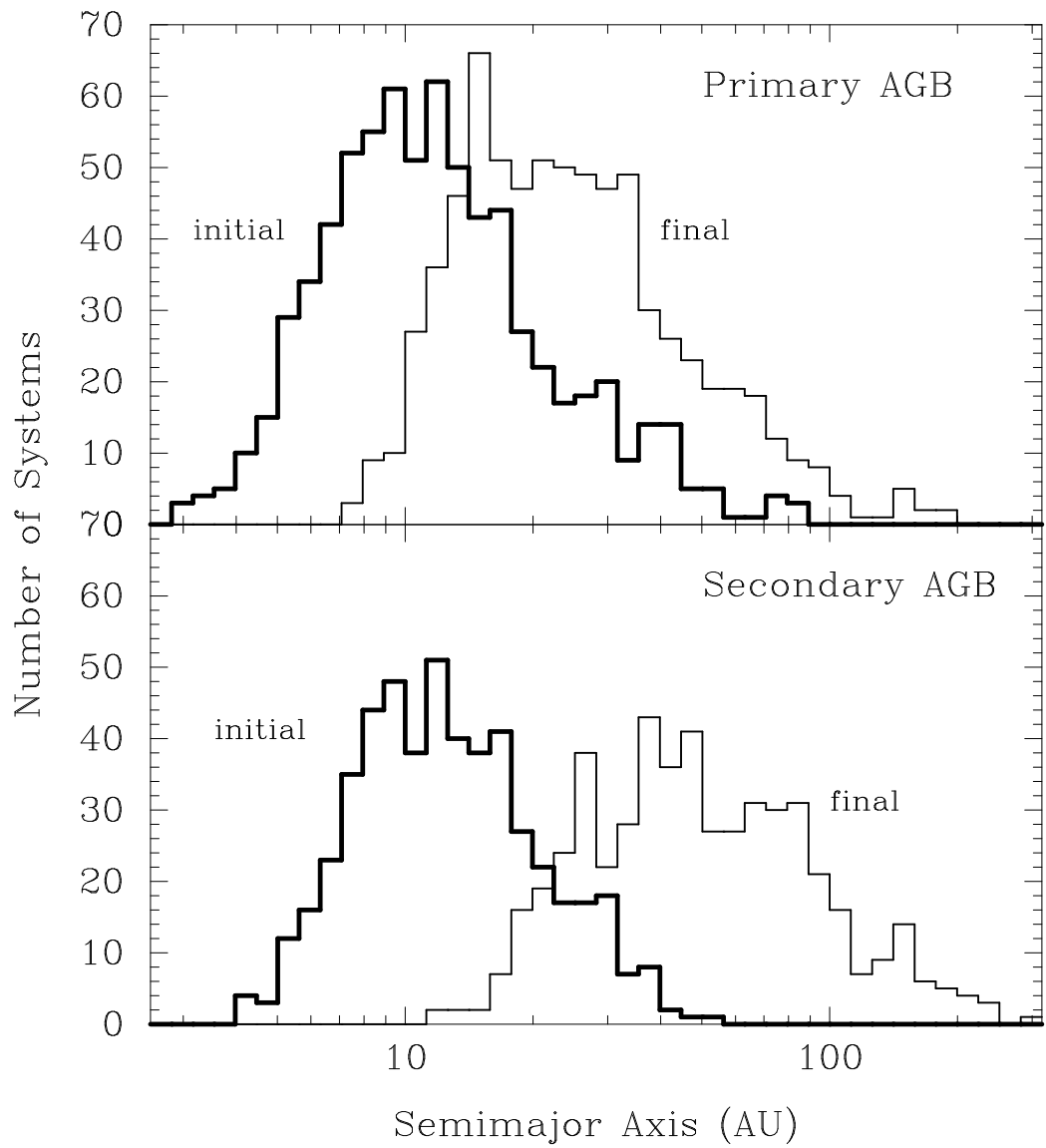
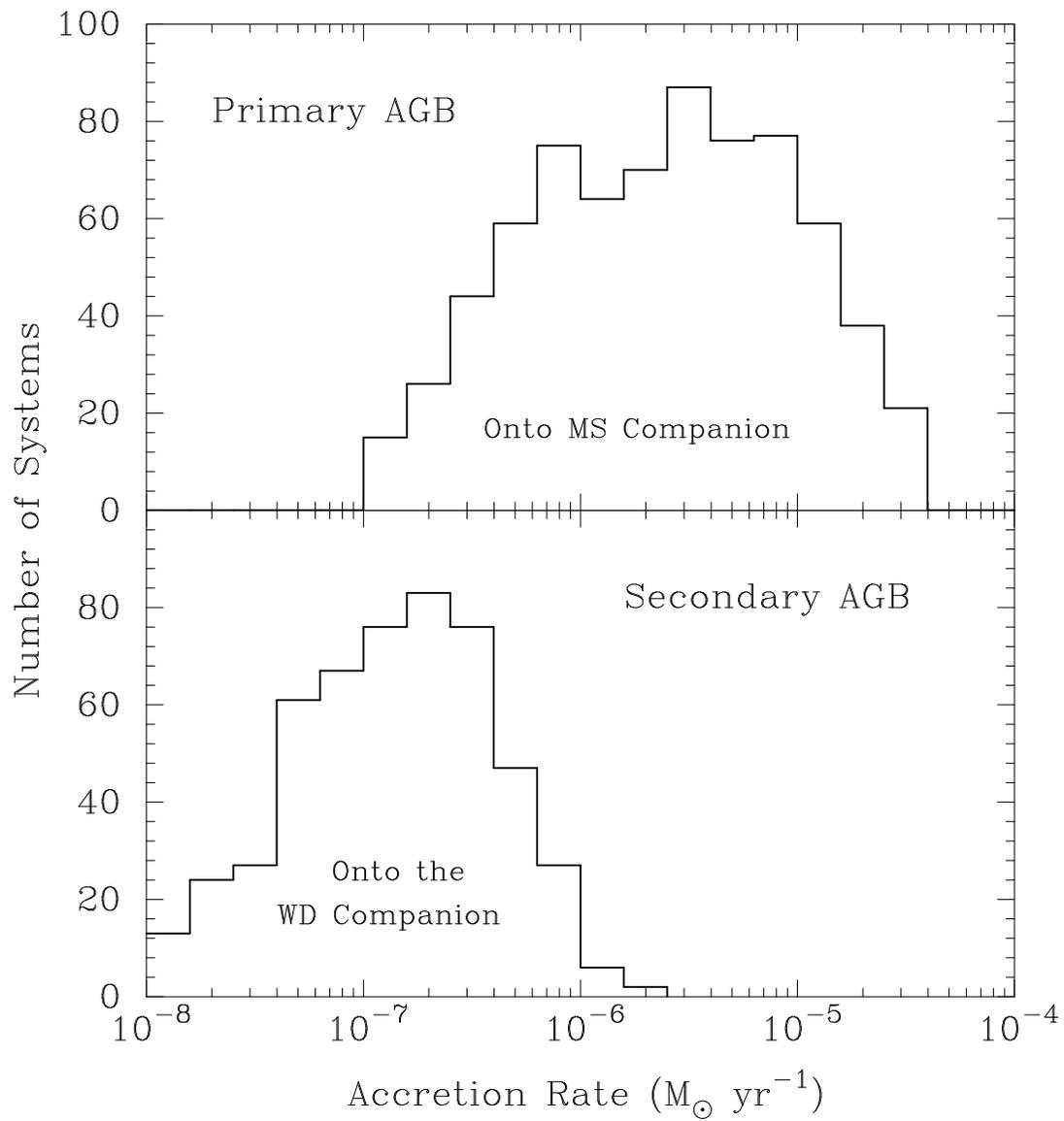
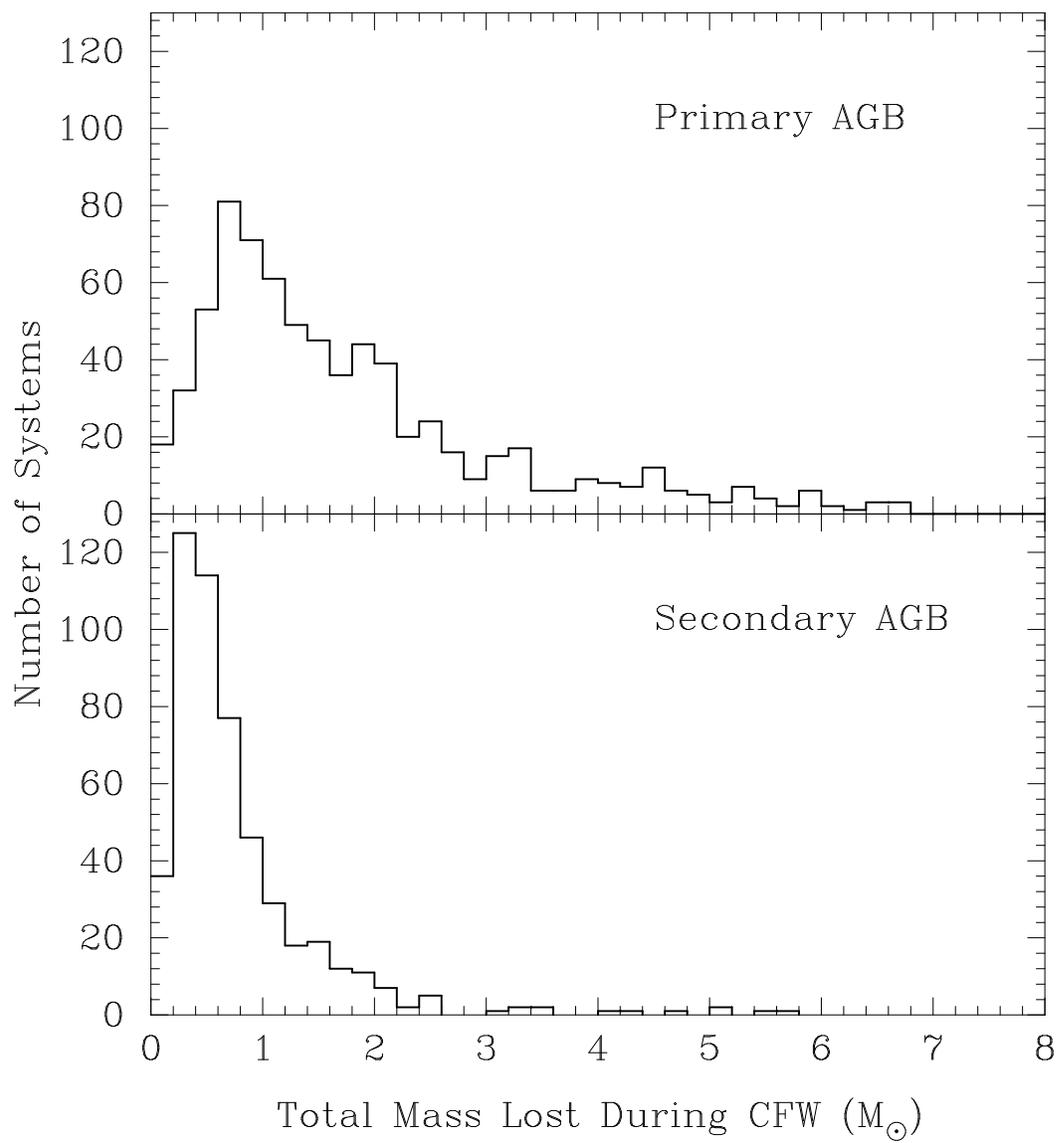
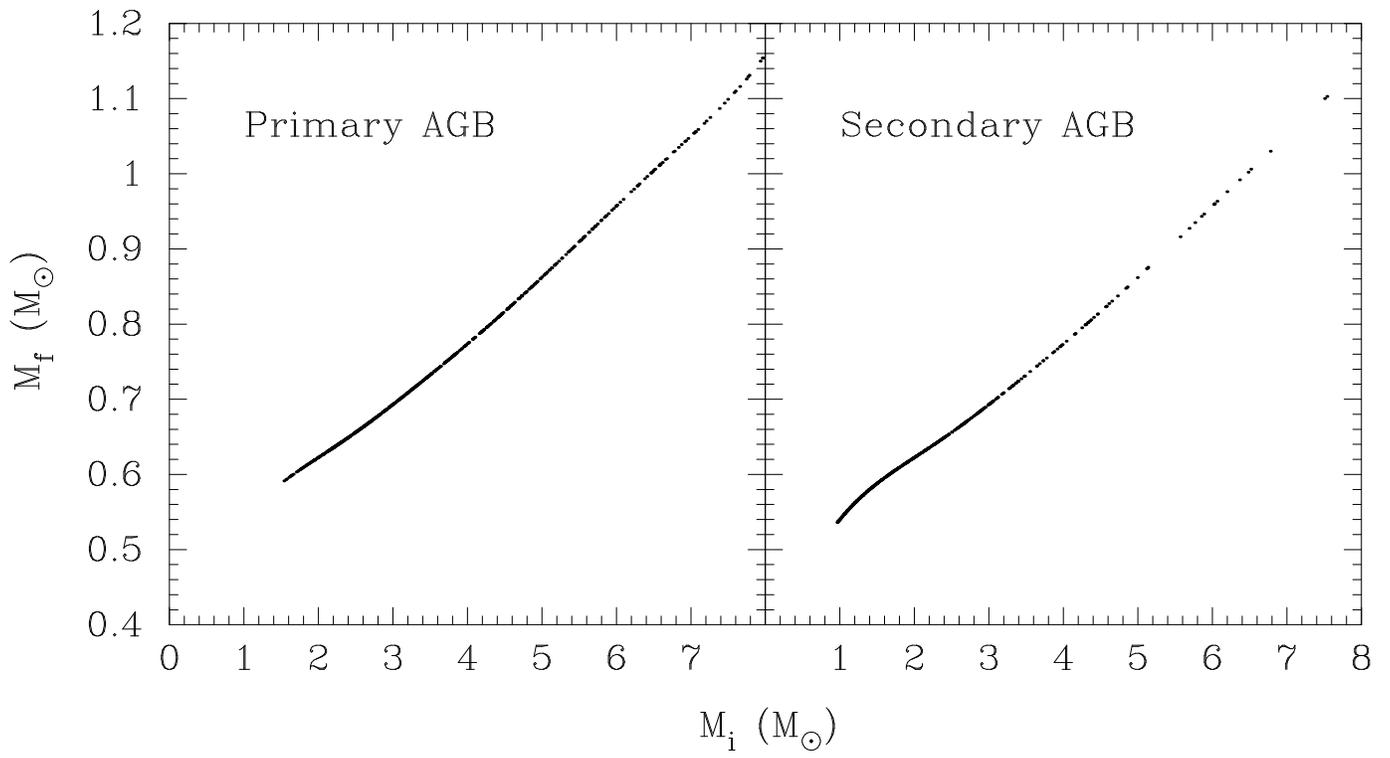
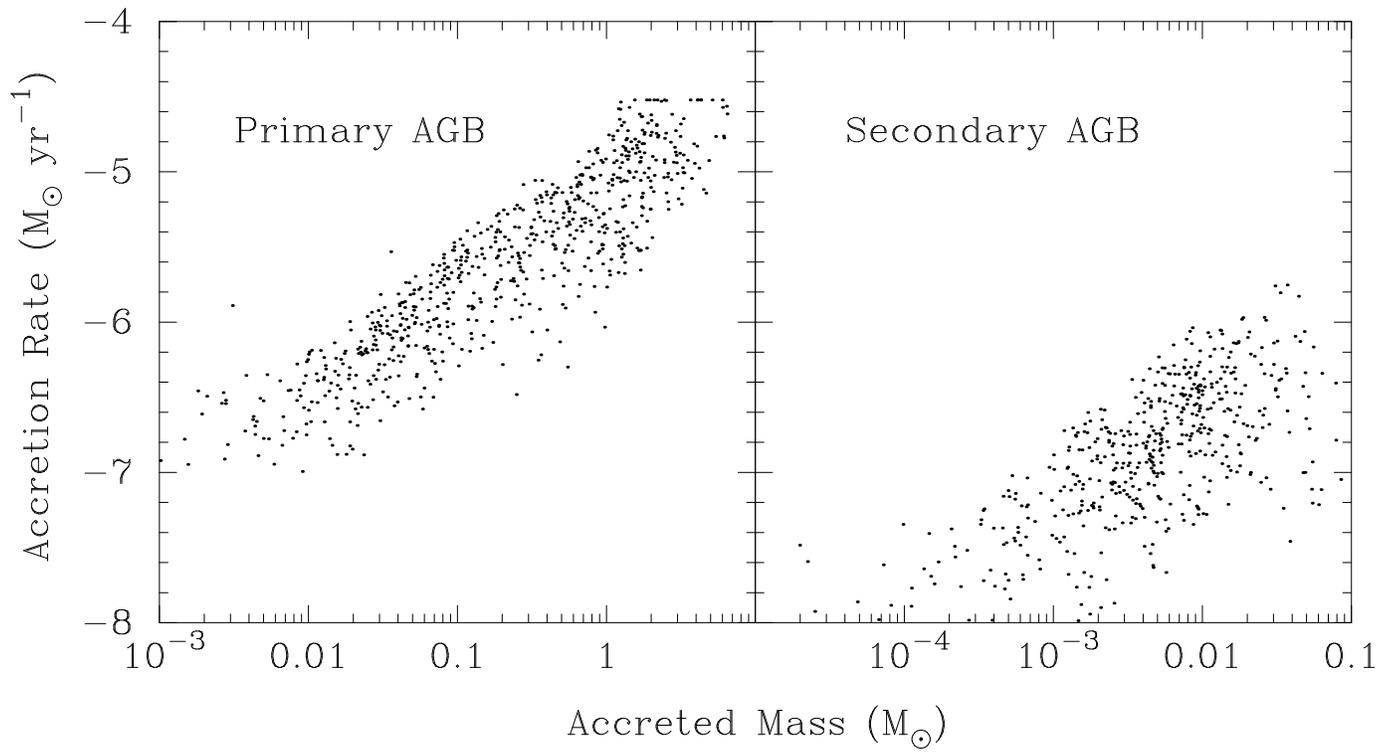
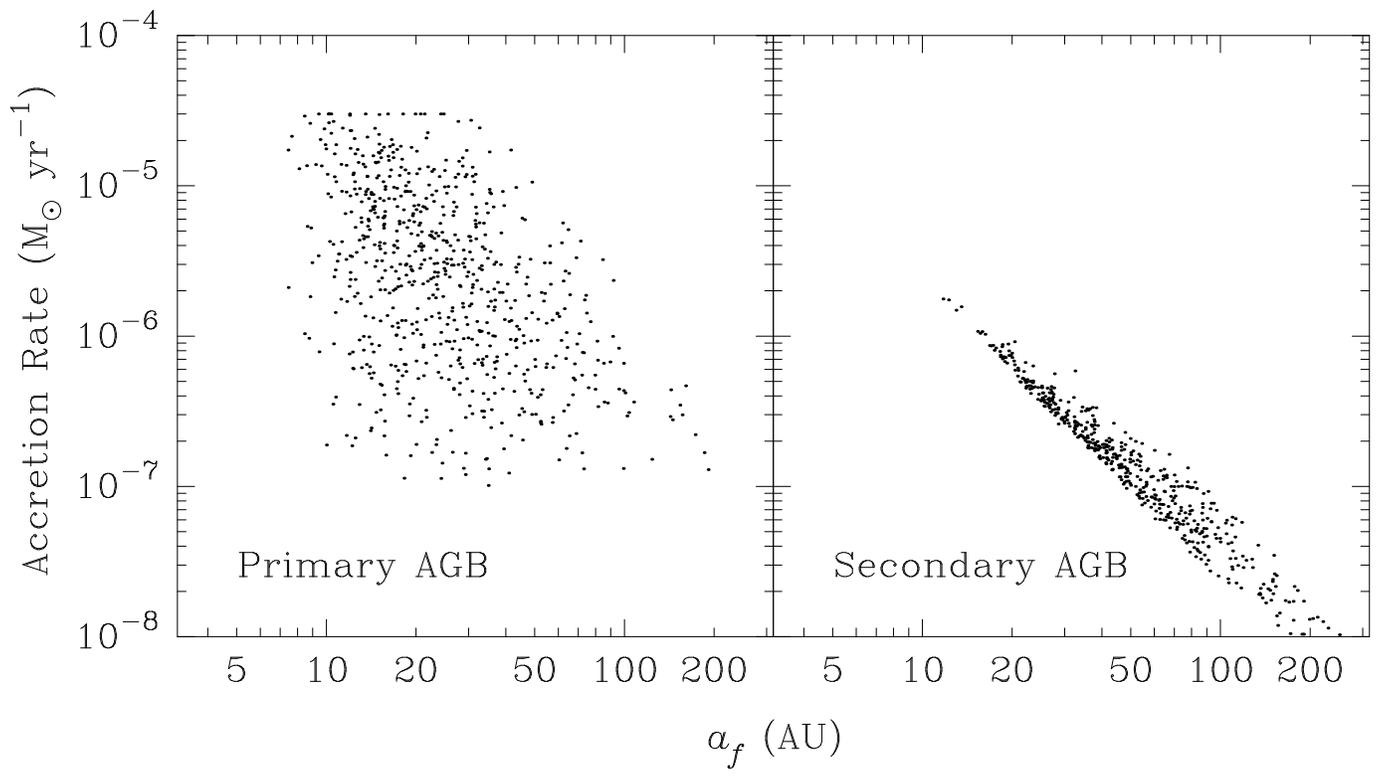
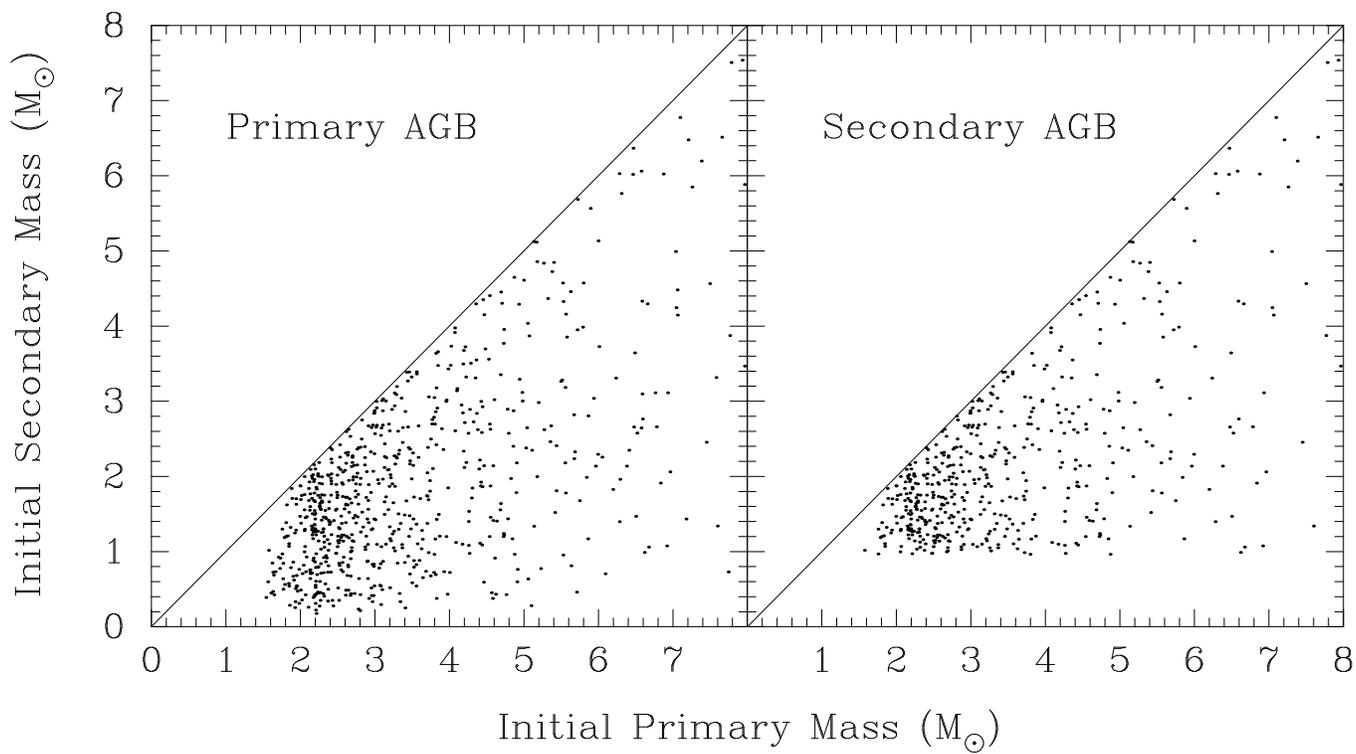
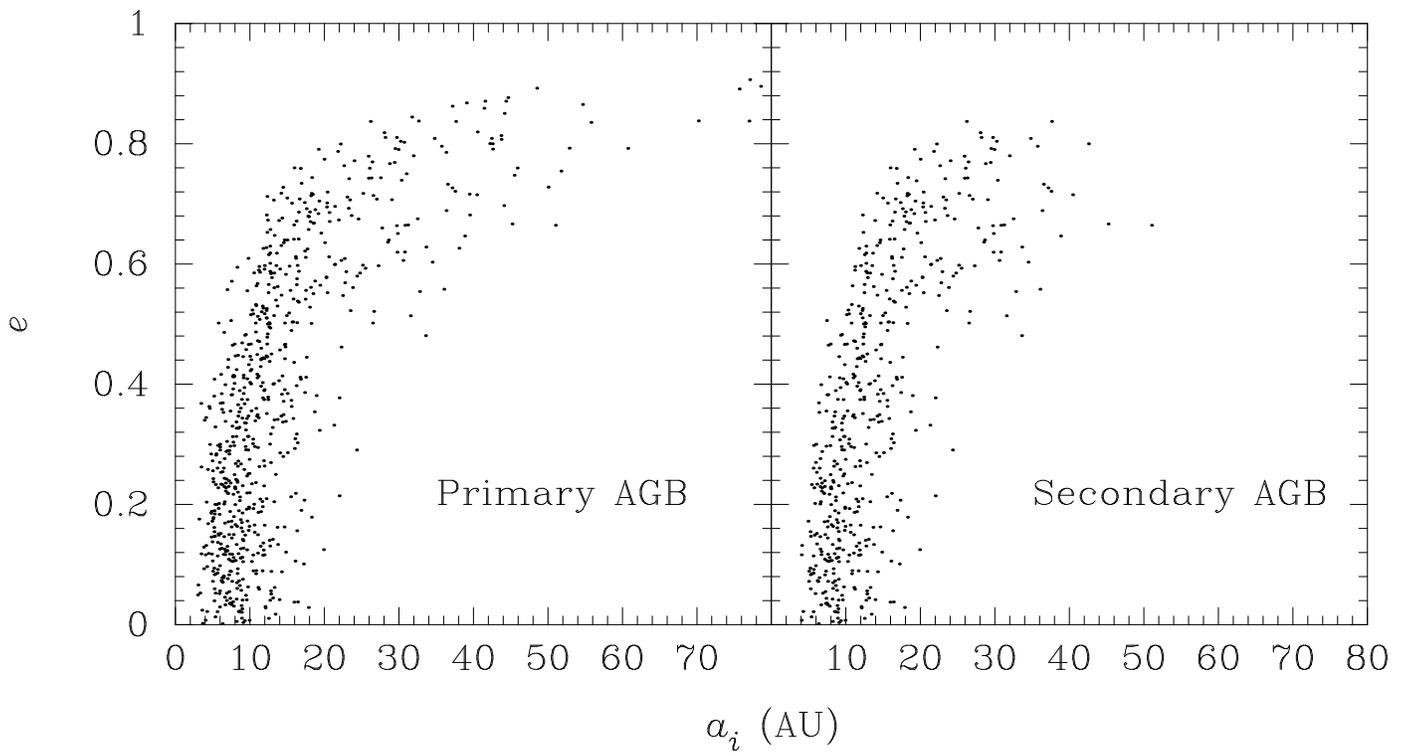
\end{document}